\documentclass[pre,aps,10pt,twocolumn,floatfix]{revtex4-1}

\usepackage[nice]{nicefrac}
\usepackage{graphicx,color}
\usepackage{amsmath,amssymb,bm,bbm}
\usepackage{amsmath}
\usepackage{braket}

\usepackage[plainpages=false,pdfpagelabels,colorlinks=true,linkcolor=red,urlcolor=blue,citecolor=blue,pdftitle={},pdfauthor={},pdfdisplaydoctitle=true,pdfduplex=DuplexFlipLongEdge]{hyperref}

\definecolor{darkred}{rgb}{0.90,0.2,0.2}
\definecolor{darkgreen}{rgb}{0,0.60,.2}
\definecolor{darkblue}{rgb}{0.1,0.3,1}
\definecolor{grey}{cmyk}{0,0,0,0.25}
\definecolor{orange}{cmyk}{0,0.6,0.8,0}

\begin{document}

\title{Tight-binding billiards}

\author{Iris Ul\v cakar}
\author{Lev Vidmar}
\affiliation{Department of Theoretical Physics, J. Stefan Institute, SI-1000 Ljubljana, Slovenia}
\affiliation{Department of Physics, Faculty of Mathematics and Physics, University of Ljubljana, SI-1000 Ljubljana, Slovenia\looseness=-1}

\begin{abstract}
Recent works have established universal entanglement properties and demonstrated validity of single-particle eigenstate thermalization in quantum-chaotic quadratic Hamiltonians.
However, a common property of all quantum-chaotic quadratic Hamiltonians studied in this context so far is the presence of random terms that act as a source of disorder.
Here we introduce tight-binding billiards in two dimensions, which are described by non-interacting spinless fermions on a disorder-free square lattice subject to curved open (hard-wall) boundaries.
We show that many properties of tight-binding billiards match those of quantum-chaotic quadratic Hamiltonians:
the average entanglement entropy of many-body eigenstates approaches the random matrix theory predictions
and one-body observables in single-particle eigenstates obey the single-particle eigenstate thermalization hypothesis.
On the other hand, a degenerate subset of single-particle eigenstates at zero energy (i.e., the zero modes) can be described as chiral particles whose wavefunctions are confined to one of the sublattices.
\end{abstract}
\maketitle

\section{Introduction}

The past three decades have established new perspectives in studies of condensed-matter lattice systems, inspired by concepts from other fields of physics such as statistical physics, random matrix theory and quantum information.
For example, the question of whether observables in isolated many-body systems in a lattice, driven away from equilibrium, agree with predictions of statistical physics ensembles after long times, has been in many ways triggered by insights in random matrix theory (RMT)~\cite{mehta_91, haake_gnutzmann_18, dalessio_kafri_16}.
As one of the nontrivial extensions of random matrix theory one usually considers the eigenstate thermalization hypothesis (ETH)~\cite{deutsch_91, srednicki_94, rigol_dunjko_08, dalessio_kafri_16}, which explains the mechanism of thermalization in generic nonintegrable systems.
In a similar way, the distinction between ground states of lattice Hamiltonians and highly excited states has benefited from results of quantum information theory.
A well known example represent bipartite entanglement entropies that may scale with the area or volume of a subsystem, thereby providing in many cases an efficient distinction between ground states of local Hamiltonians~\cite{eisert_cramer_10} and excited eigenstates~\cite{bianchi_hackl_22}, respectively.

\begin{figure}[b]
\centering
\includegraphics[width=0.98\columnwidth]{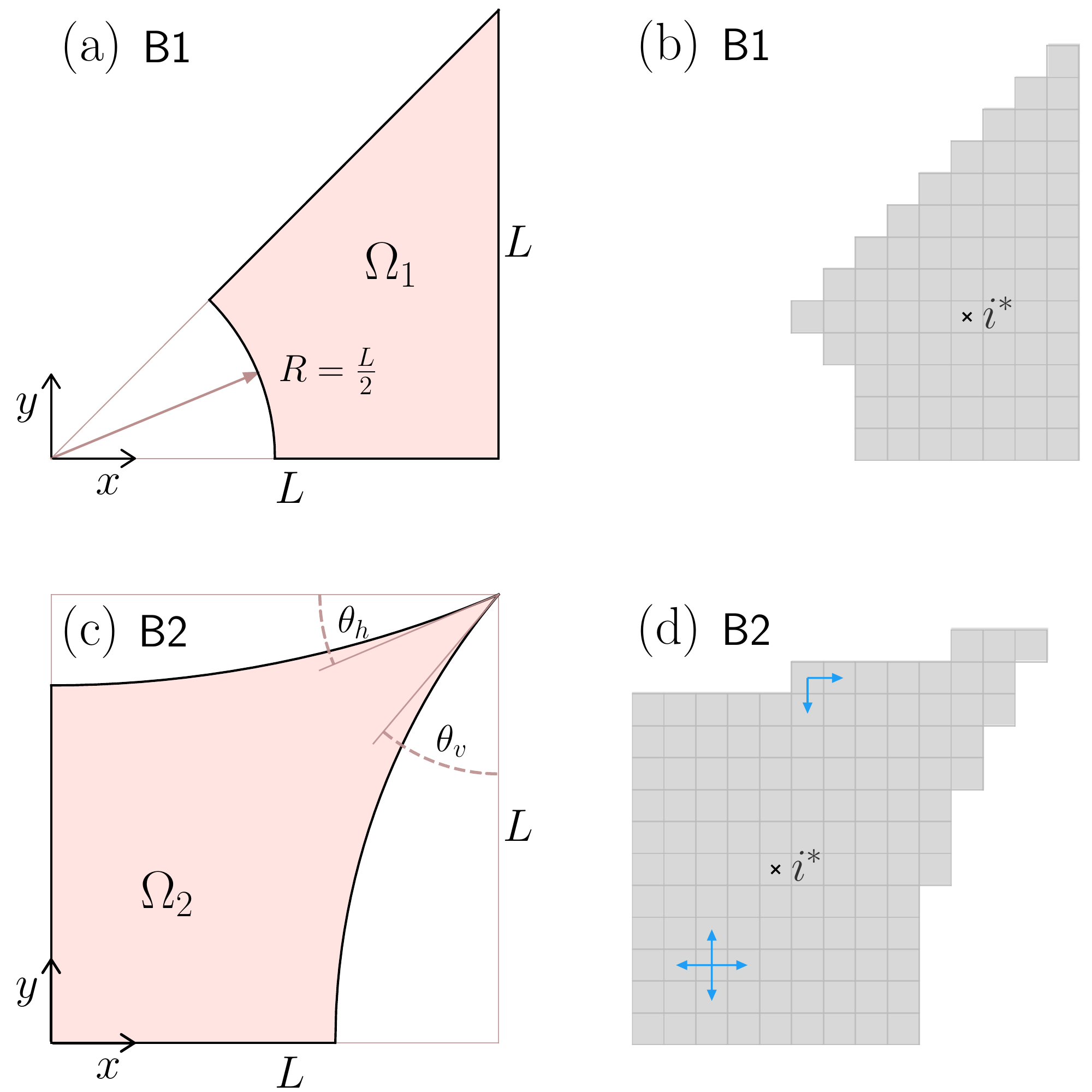}
\caption{
Tight-binding billiards studied in this work.
(a, b) Billiard 1 (B1). (c, d) Billiard 2 (B2).
Left column shows the boundaries in continuum given by Eqs.~(\ref{eq:Sinai}) and~(\ref{eq:Loka}), while the right column shows examples of the lattices used in actual calculations (a lattice site is in a center of a square).
The ''center of mass'' is denoted by $i^*$.
}\label{fig1}
\end{figure}

An important class of Hamiltonians that are the focus of this work are quadratic Hamiltonians, which can be expressed by bilinear forms in creation and annihilation operators.
Several quadratic Hamiltonians, such as the three-dimensional Anderson model below the localization transition and the quadratic Sachdev-Ye-Kitaev (SYK2) model, exhibit statistical properties of single-particle energy spectrum, as well as entanglement properties of many-body energy eigenstates, that are consistent with predictions of RMT~\cite{altshuler_shklovskii_86, altshuler_zharekeshev_88, shklovskii_shapiro_93, hofstetter_schreiber_93, sierant_delande_20, suntajs_prosen_21, liu_chen_18, lydzba_rigol_20, lydzba_rigol_21}.
Moreover, properties of observables in single-particle energy eigenstates comply with the single-particle ETH~\cite{lydzba_zhang_21}.
One may hence argue that they exhibit single-particle quantum chaos.
We refer to quadratic Hamiltonians with these properties as quantum-chaotic quadratic Hamiltonians~\cite{lydzba_rigol_21}.
However, a common feature of the latter Hamiltonians studied so far is the presence of random terms that act as a source of disorder.
Here we study noninteracting spinless fermions in square lattices with open (hard-wall) boundaries that are shaped such that in the thermodynamic limit they represent walls of a billiard, see Fig.~\ref{fig1}.
We dub these systems {\it tight-binding billiards}.
Since noninteracting fermions in square lattices with regular boundaries (e.g., a rectangle) do not exhibit any signatures of quantum chaos, the chaotic behavior emerges due to a particular shape of the open boundaries.
Understanding to what extent do the tight-binding billiards of noninteracting fermions comply with single-particle quantum chaos is the main motivation of our work.

Our work is in part also motivated by experimental activities in quantum nanostructures~\cite{ravnik_vaskivskyi_21}.
Applying laser-induced quenches to a TaS$_2$ material, the authors of Ref.~\cite{ravnik_vaskivskyi_21} recently managed to create geometrical confinement with atomically precise shapes that resemble lattice billiards.
While our study does not attempt to mimic the actual experimental conditions, it aims to shed some light onto the statistical properties and the entanglement content in these systems, and paves the way towards further applications.
Other lattice systems that carry certain similarities to the tight-binding billiards studied here include, e.g., free particles on square and cubic lattices with disorder on the boundaries~\cite{pavloff_hansen_92, cuevas_louis_96}, graphene lattices with confinement~\cite{libisch_stampfer_09, wimmer_akhmerov_10, huang_lai_10, hunag_xu_14, yu_li_16, hagymasi_vancso_17}, and interacting two-particle systems~\cite{perrin_asboth_21, poshakinskiy_zhong_21}.

We note that the tight-binding billiards studied here may be seen as a lattice version of quantum billiards in continuum, for which the dynamics is expected to be fully chaotic and their spectral properties comply with the RMT predictions~\cite{bohigas_giannoni_84, rudnick_08}.
Studies of quantum billiards in continuum such as those in stadium~\cite{mcdonald_kaufman_79, casati_valzgris_80} or Sinai~\cite{berry_81, bohigas_giannoni_84} billiards played a major role in early works of single-particle quantum chaos~\cite{guhr_muller_98, stoeckmann_99}, and introduced several important concepts such as the quantum chaos conjecture~\cite{bohigas_giannoni_84}.
However, when the lattice constant $a$ is sent to zero, tight-binding billiards correspond to an ultra-high energy regime of continuous billiards, and have therefore not received much attention in previous studies.
Consider, e.g., a typical energy $E \sim k^2$ in a continuum, where $k$ is a wave number.
If one then introduces a typical energy scale of the system in a lattice, where $ka = O(1)$, this results in $E \sim (ka)^2/a^2$, i.e., the typical energy scale of tight-binding billiards diverges as $\propto 1/a^2$.
This is a much larger scale than the semi-classical scale $\propto 1/l^2$ in continuous billiards, where $l$ is the smallest length scale describing the continuous boundary geometry.
The lattice discretization hence prohibits a straight-forward connection of our results to billiards in continuum, and we refrain from making any quantitative comparison between these two regimes.

We study three main properties of tight-binding billiards of noninteracting fermions from Fig.~\ref{fig1}:
(a) statistics of nearest level spacings of single-particle energy spectrum,
(b) von Neumann and 2nd R\' enyi bipartite eigenstate entanglement entropies, averaged over many-body eigenstates, and
(c) properties of matrix elements of local observables in single-particle eigenstates.
Our main results is that these properties indeed appear to be consistent with the emergence of quantum chaos in tight-binding billiards.
This statement nevertheless needs to be taken with some care since there exist a sub-extensive (in lattice volume) set of single-particle eigenstates that are degenerate in the middle of the spectrum at zero energy (i.e., zero modes), for which the agreement with RMT predictions may not be established.
We describe several key properties of zero modes and argue that they are eigenstates of a chiral operator, which gives rise to a confinement of their wavefunction amplitudes to one of the sublattices.

Focusing on energy eigenstates away from zero modes, we find that the level spacing statistics agree with RMT predictions to extremely high accuracy, while the average volume-law eigenstate entanglement entropies exhibit more pronounced finite-size effects.
Still, the volume-law contribution to the entanglement entropy appears to be well described with recent predictions for quantum-chaotic quadratic systems based on the RMT analysis~\cite{lydzba_rigol_20, lydzba_rigol_21}.
As a side result, we extend the latter predictions to derive a closed form expression for the 2nd R\' enyi entanglement entropy.
Finally, we find that local observables in single-particle eigenstates comply with the single-particle ETH, and show that the distributions of both diagonal and off-diagonal matrix elements are in general not Gaussian, as recently argued in Ref.~\cite{lydzba_zhang_21}.

The paper is organized as follows.
In Sec.~\ref{sec:models} we introduce the model and lattice geometries under consideration, and in Sec.~\ref{sec:zeromodes} we discuss some properties of zero modes.
We study statistical properties of the single-particle spectrum in Sec.~\ref{sec:spectrum}, while in Sec.~\ref{sec:entanglement} we turn our attention to entanglement properties of many-body eigenstates.
In Sec.~\ref{sec:observables} we study statistical properties on matrix elements of local observables in single-particle eigenstates.
We conclude in Sec.~\ref{sec:conclusions}.

\section{General considerations} \label{sec:general}

\subsection{Models and geometry of tight-binding billiards} \label{sec:models}

We study the tight-binding Hamiltonian on a square lattice,
\begin{equation}
\label{eq:billiard}
    \hat{H} = -t\sum_{\substack{\langle i, j\rangle \in \Omega}}\left(\hat{c}_i^{\dagger}\hat{c}_j + \hat{c}_j^{\dagger}\hat{c}_i\right),
\end{equation}
where $\hat{c}_i^{\dagger}$ and $\hat{c}_i$ are creation and annihilation operators of spinless fermions on site $i$, respectively, and the sum runs over nearest neighbors.
We set $t\equiv 1$ further on.
The lattice consists of $V$ sites, also referred to as the lattice volume.
The lattice sites belong to a closed region denoted by $\Omega$, which represents a discretized representation of the continuous billiards introduced below
(for simplicity, we use the label $\Omega$ in both discrete and continuous space).
We use open (hard-wall) boundary conditions, i.e., hopping on sites outside the region $\Omega$ is forbidden.

The definition of $\Omega$ proceeds in four steps.
(i)
We define a large square of size $L \times L$ and a coordinate system $(x,y)$ whose origin is in the bottom left corner of a square, as sketched in Figs.~\ref{fig1}(a) and~\ref{fig1}(c).
(ii)
We split the large square into $L^2$ equal squares with unit sides.
We introduce a square lattice for which the sites are located in the center of each unit square.
A site labeled  $i$ has then spatial coordinates $i_x$, $i_y$.
(iii)
We define boundaries of a continuous billiard.
The boundaries of specific billiards studied in this work are introduced in the next paragraph.
(iv)
At a given $L$, the sites of a lattice whose coordinates are inside the boundaries of a continuous billiard belong to the region $\Omega$ on which the Hamiltonian in Eq.~(\ref{eq:billiard}) is defined.
We verify that each lattice site in $\Omega$ is connected to at least one neighboring site.

We consider two billiards, which are in the literature usually referred to as variants of Sinai billiards.
The first, referred to as billiard 1 (shortly, B1), is an isosceles orthogonal triangle with legs of size $L$ and a circular cut at one of the corners with angle $\pi/4$, as shown in Fig.~\ref{fig1}(a).
The radius of the circle is $R=rL$.
By default we only consider $r=1/2$ as in Fig.~\ref{fig1}(a), while in some cases we complement the results at $r=1/2$ with those at $r=1/3$ (in general, we do not observe any qualitative differences between these two realizations of the billiard B1).
The region of the billiard is defined as
\begin{equation}
\label{eq:Sinai}
    \Omega_1 = \left\{(x, y)\,\Big\vert\; 0 \leq y \leq x \leq L \;\wedge\; x^2 + y^2 \geq R^2\right\}.
\end{equation}
The second, referred to as billiard 2 (shortly, B2), consists of a square that is cut by two circular arcs, as shown in Fig.~\ref{fig1}(c).
The arcs intercept in the upper right corner of the square, the first arc forming an angle $\theta_{\rm h} = 0.4$ with the horizontal axis and the second arc an angle $\theta_{\rm v} = 0.7$ with the vertical axis.
The circular arcs are formed by circles with the radii $R_{\rm h} \approx 2.57L$ and $R_{\rm v} \approx 1.55L$, and the centers $(x_{\rm h}, y_{\rm h}) \approx (0, 3.37L)$ and $(x_{\rm v}, y_{\rm v}) \approx (2.19L, 0)$, respectively.
The region of the billiard is defined as
\begin{equation}
\label{eq:Loka}
\begin{aligned}[b]
    \Omega_2 & = \Big\{(x, y) \,\Big\vert\; 0 \leq x \leq L \;\wedge\; 0 \leq y \leq L \;\wedge\; \\
    & x^2 + (y-y_{\rm h})^2 \geq R_{\rm h}^2 \;\wedge\; (x-x_{\rm v})^2 + y^2 \geq R_{\rm v}^2\Big\} \;.
\end{aligned}
\end{equation}
The regions $\Omega_1$ and $\Omega_2$ from Eqs.~(\ref{eq:Sinai}) and~(\ref{eq:Loka}) are then used to define the tight-binding billiards in a lattice, as described in the previous paragraph.
Figures~\ref{fig1}(b) and~\ref{fig1}(d) show the tight-binding billiards B1 and B2 at $L = 14$ with the total number of lattice sites $V_1 = 74$ and $V_2 = 119$, respectively.
Blue arrows show examples of the allowed hoppings from selected lattice sites (bases of the arrows) to their neighbouring sites (tips of the arrows). 

\begin{figure}[!t]
\centering
\includegraphics[width=0.98\columnwidth]{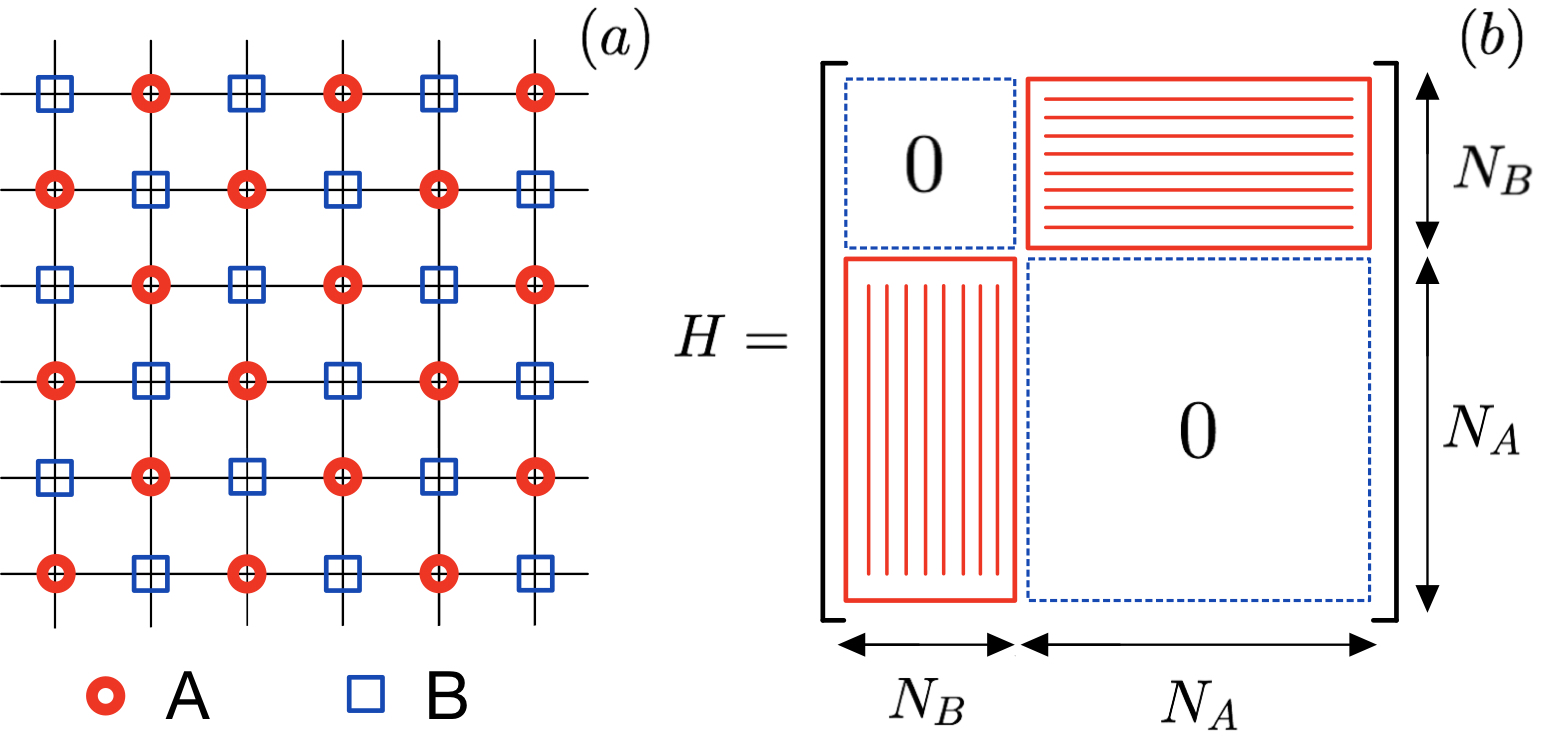}
\caption{
(a) A segment of a lattice of tight-binding billiards, far away from the boundaries.
The Hamiltonian $\hat H$ in Eq.~(\ref{eq:billiard}) acts such that a particle from the sublattice A hops into the sublattice B, and vise versa.
However, due to curved boundaries, tight-binding billiards need not contain the same number of lattice sites in sublattices A and B.
(b) A sketch of the Hamiltonian matrix in case when the number $N_A$ of sites in sublattice A is larger than the number $N_B$ of sites in sublattice B.
Red stripes denote the region of the matrix where the matrix elements may be nonzero.
}\label{fig_sublattice}
\end{figure}

The billiards that we study do not exhibit any spatial symmetries, apart from the chiral (sublattice) symmetry that is discussed in Sec.~\ref{sec:zeromodes}.
The single-particle energy spectrum of the Hamiltonian $\hat H$ from Eq.~(\ref{eq:billiard}) consists of two types of eigenstates.
The first are eigenstates with nonzero energy, $|E_\alpha| > 0$.
These always occur in pairs of positive and negative energy, i.e., for a nonzero energy $E_\alpha$ there exist another eigenstate with energy $E_\beta$, such that $E_\beta = - E_\alpha$.
The origin of this reflection symmetry of the single-particle energy spectrum is explained below.
The second subset of eigenstates are those with energy $E_\alpha=0$, denoted as zero modes.
We study the latter in more details in Sec.~\ref{sec:zeromodes}, where we also show that the number of zero modes increases with $V$, however, the fraction of zero modes relative to all single-particle eigenstates vanishes in the thermodynamic limit $V\to\infty$.

The reflection symmetry of the energy levels is a consequence of the bipartite nature of the Hamiltonian $\hat H$ from Eq.~(\ref{eq:billiard}).
With the latter we have in mind a division of the lattice into sublattices A and B, see Fig.~\ref{fig_sublattice}(a), such that the action of $\hat H$ on a particle in sublattice A moves the particle into sublattice B, and vice versa.
Then, one can introduce an operator $\hat \Gamma$, whose action on an arbitrary superposition of single-particle states from both sublattices is of the form
\begin{equation} \label{def_gamma}
    \hat \Gamma \left(\hat c_{i_A}^\dagger|\emptyset\rangle + \hat c_{i_B}^\dagger|\emptyset\rangle\right) = c_{i_A}^\dagger|\emptyset\rangle - \hat c_{i_B}^\dagger|\emptyset\rangle \;,
\end{equation}
where $\hat c_{i_A}^\dagger$ and $\hat c_{i_B}^\dagger$ create spinless fermions in sublattices A and B, respectively.
(One may equivalently consider the operator $-\hat \Gamma$, with identical conclusions.)
Equation~(\ref{def_gamma}) suggests that $\hat \Gamma^2 = \hat I$, and hence its eigenvalues are $\pm 1$.
The reflection symmetry of the energy levels originates from the anticommutation relation
\begin{equation} \label{def_anticommutator}
\{\hat \Gamma, \hat H\} = \hat \Gamma \hat H + \hat H \hat \Gamma = 0 \;.
\end{equation}
Consequently, $\hat \Gamma$ does not represent a true symmetry of the model, however, we show in Sec.~\ref{sec:zeromodes} that zero modes can be made eigenstates of $\hat \Gamma$.
Operators with properties similar to those of $\hat \Gamma$ are commonly associated with chiral or sublattice symmetry~\cite{ryu_schnyder_10, ludwig_15, zirnbauer_21}, and we refer to it as a chiral operator further on.
In terms of spinless fermion creation and annihilation operators, $\hat \Gamma$ acts as
\begin{equation}
    \hat \Gamma \hat c_i^\dagger \hat \Gamma = \theta_i \hat c_i^\dagger \;,\;\;\;
    \hat \Gamma \hat c_i \hat \Gamma = \theta_i \hat c_i \;,
\end{equation}
where $\theta_i=+1$ for site $i$ in sublattice A and $\theta=-1$ for site $i$ in sublattice B.

Several simple properties follow from Eq.~(\ref{def_anticommutator}).
Let us express a single-particle eigenstate $|\alpha\rangle$ of $\hat H$ as
\begin{equation} \label{def_alpha}
    |\alpha\rangle = \sum_{i\in A} u_i^{(\alpha)} |i\rangle + \sum_{j\in B} v_j^{(\alpha)} |j\rangle \;,
\end{equation}
where the wavefunction coefficients $u_i^{(\alpha)}$ and $v_j^{(\alpha)}$ correspond to single-particle states in sublattices A and B, respectively, and $|i\rangle \equiv \hat c_{i}^\dagger |\emptyset\rangle$, $|j\rangle \equiv \hat c_{j}^\dagger |\emptyset\rangle$.
As a consequence of Eq.~(\ref{def_anticommutator}), there exists a state $\hat \Gamma |\alpha\rangle$ that is also an eigenstate of $\hat H$, with the property
\begin{equation} \label{def_H_Gamma_alpha}
    \hat H \left( \hat \Gamma|\alpha\rangle \right) = - E_\alpha \left( \hat \Gamma|\alpha\rangle \right) \;.
\end{equation}
Defining $|\bar\alpha\rangle \equiv \hat \Gamma |\alpha\rangle$, one observes that the wavefunction of $|\bar\alpha\rangle$ can be expressed using identical weights as the one of $|\alpha\rangle$, however with alternating signs,
\begin{equation} \label{def_beta}
    |\bar\alpha\rangle = \sum_{i\in A} u_i^{(\alpha)} |i\rangle - \sum_{j\in B} v_j^{(\alpha)} |j\rangle \;.
\end{equation}
This result suggests that it may be sufficient for certain properties to consider only the lower part of the spectrum, $E_\alpha \leq 0$.
Moreover, due to orthogonality of the wavefunctions $|\alpha\rangle$ and $|\bar\alpha\rangle$ it follows that
\begin{equation}
    \sum_{i\in A} \left(u_i^{(\alpha)}\right)^2 = \sum_{j\in B} \left(v_j^{(\alpha)}\right)^2 = \frac{1}{2} \;
\end{equation}
for any eigenstate with $E_\alpha \neq 0$, which can be seen as an additional sum rule defined for each of the sublattices.

We complement our analysis by considering tight-binding billiards with additional onsite potentials, which break the bipartite nature of the Hamiltonian $\hat H$ from Eq.~(\ref{eq:billiard}).
To this end we introduce two types of Hamiltonians.
In the first case we add a harmonic potential,
\begin{equation}
\label{eq:billiard_ho}
    \hat{H}_k = \hat{H} + k \sum_{i\in\Omega} \left(\frac{\left(i_x-i_x^*\right)^2 + \left( i_y-i_y^*\right)^2}{k_{\rm max}}\right)\hat{c}_i^{\dagger}\hat{c}_i \;,
\end{equation}
where $k$ is the strength of the potential, $i_x$, $i_y$ are the $x$, $y$ spatial coordinates of site $i$, and $i^*_x$, $i^*_y$ are the coordinates of $i^*$ that is close to the lattice center of mass, as depicted in Figs.~\ref{fig1}(b) and~\ref{fig1}(d).
In our implementation we also use a normalization number $k_{\text{max}}$, which is set such that the largest potential (i.e., the potential at the site that is most distant from the center of mass) equals $k$.
In the second case we add random potentials on each lattice site,
\begin{equation}
\label{eq:billiard_random}
    \hat{H}_{\rm rand} = \hat{H} + \frac{W}{2}\sum_{i\in\Omega} h_i \hat{c}_i^{\dagger}\hat{c}_i \;,
\end{equation}
where $h_i$ are independent and identically distributed random variables with a uniform probability in the interval $h_i \in \left[-1, 1\right]$.
For conventional boundary conditions this model is referred to as the two-dimensional Anderson model.
The single-particle energy spectra of the Hamiltonians in Eqs.~(\ref{eq:billiard_random}) and~(\ref{eq:billiard_ho}) are not reflection-symmetric and they do not exhibit zero modes, i.e., they are nondegenerate.

\subsection{Zero modes and chiral symmetry} \label{sec:zeromodes}

Both tight-binding billiards under considerations, described by $\hat H$ from Eq.~(\ref{eq:billiard}), exhibit a large number of zero modes.
Below we show that a sufficient condition for the emergence of zero modes requires two properties:
existence of a chiral operator from Eq.~(\ref{def_gamma}) that anticommutes with $\hat H$, and a nonzero sublattice imbalance $\delta N$ between sites on sublattices A and B, see Eq.~(\ref{def_delta_N}).

The Hamiltonian $\hat H$ from Eq.~(\ref{eq:billiard}) can be represented by a matrix whose indices first run over sites in one sublattice and then over its complement.
Such a matrix consists of four blocks, two diagonal and two off-diagonal blocks, as sketched in Fig.~\ref{fig_sublattice}(b).
The dimensions of the diagonal blocks are $N_A \times N_A$ and $N_B \times N_B$, where $N_A$ and $N_B$ denote the number of lattice sites in sublattices A and B, respectively, and $N_A+N_B=V$.
As a consequence of anticommutation of $\hat \Gamma$ with $\hat H$, Eq.~(\ref{def_anticommutator}), the diagonal blocks are zero.
In contrast, the off-diagonal blocks of dimensions $N_A \times N_B$ and $N_B \times N_A$ may include nonzero elements.
We assume that $N_A \geq N_B$, as in Fig.~\ref{fig_sublattice}(b), and we define the sublattice imbalance $\delta N$ as
\begin{equation} \label{def_delta_N}
    \delta N = N_A - N_B\;.
\end{equation}
In what follows we show that the total number of zero modes, denoted as $\cal M$, is at least $\delta N$.

Let us assume that the number of linearly independent columns in the left block of $H$ in Fig.~\ref{fig_sublattice}(b) [i.e., column indices $l=1,...,N_B$] is $N_B - m$, where $m \geq 0$ is a non-negative integer.
Then, the number of linearly independent rows (and therefore columns) in the right block of $H$ [i.e., column indices $l=N_B+1,...,N_B+N_A$] is also $N_B - m$ since $H$ is hermitian.
The rank of the full Hamiltonian matrix $H$ is therefore
\begin{equation} \label{def_rank_H_m}
    {\rm rank}(H) = N_B - m + N_B - m = V - (\delta N + 2m) \;,
\end{equation}
and the number of zero modes is
\begin{equation} \label{def_M_exact}
    {\cal M} = \delta N + 2m \;.
\end{equation}
Since $m$ can not be negative, Eq.~(\ref{def_M_exact}) suggests that the lower bound on $\cal M$ is $\delta N$.

\begin{figure}[!t]
\centering
\includegraphics[width=0.98\columnwidth]{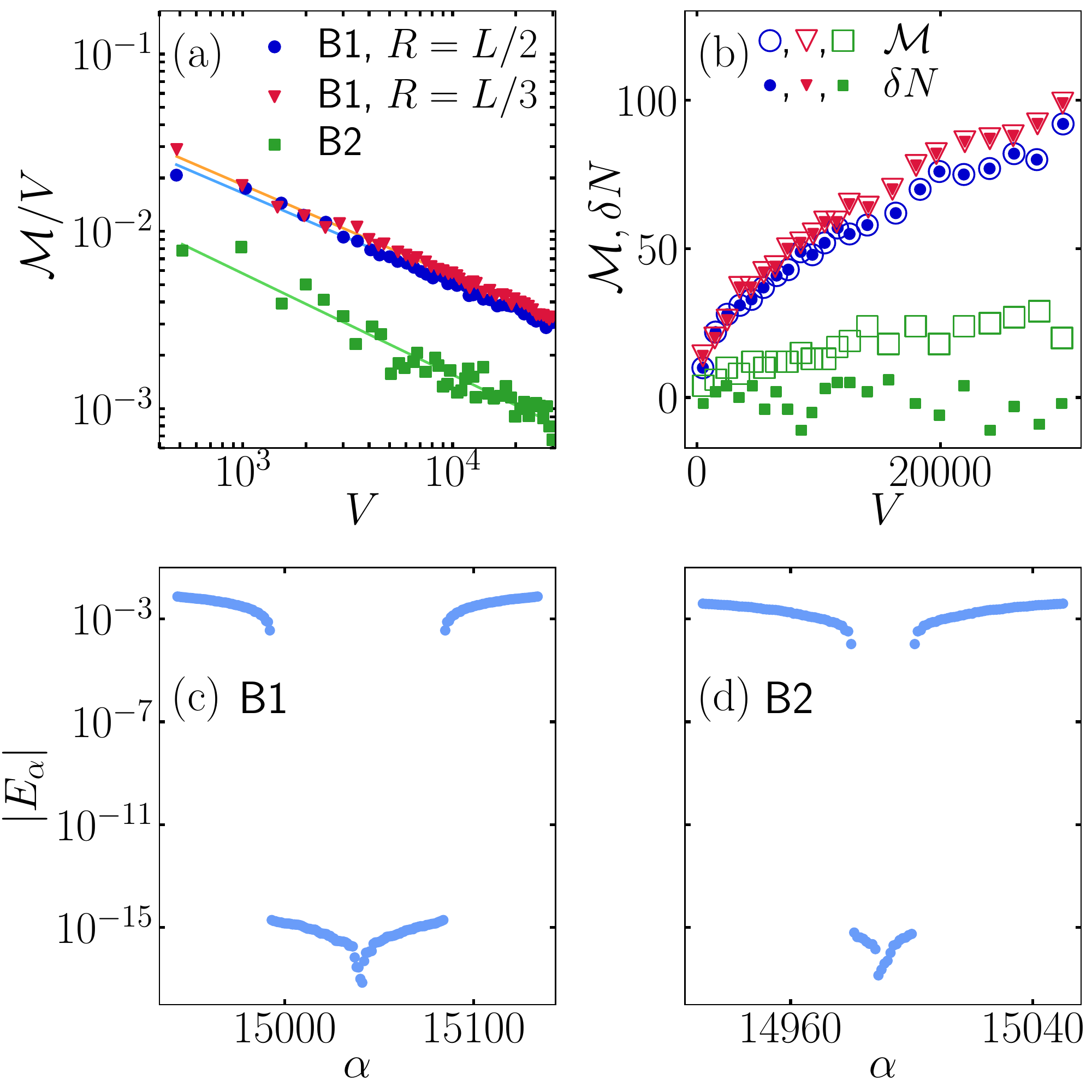}
\caption{
(a)
Fraction of zero modes ${\cal M}/V$ versus $V$.
Solid lines are fits of the function $a V^{-\zeta}$ to the results.
We obtain $\zeta=0.51$ in the billiard B1 at $R=L/2$ and $R=L/3$, and $\zeta=0.57$ in the billiard B2.
(b)
The number of zero modes ${\cal M}$ [open symbols] and the sublattice imbalance $\delta N$ from Eq.~(\ref{def_delta_N}) [filled symbols] versus $V$, for the same systems as in panel (a).
(c) and (d) 
Absolute values of single-particle eigenenergies $|E_\alpha|$ versus the eigenstate index $\alpha$ in the billiards B1 at $V = 30078$ ($R=L/2$) and B2 at $V=29982$, respectively.
We show results for $\alpha$ around the middle of the spectrum where the zero modes emerge.
We define zero modes numerically as eigenstates with energy $|E_\alpha| \leq 10^{-14}$.
}\label{fig2_2}
\end{figure}

The above arguments are corroborated by a numerical analysis in the tight-binding billiards B1 and B2 in Fig.~\ref{fig2_2}.
Figure~\ref{fig2_2}(a) shows the scaling of the total number of zero modes ${\cal M}$ relative to the total number of states $V$.
We observe in all cases that ${\cal M}/V \propto V^{-\zeta}$, where $0<\zeta <1$.
This suggests that the fraction of zero modes vanishes in the thermodynamic limit $V \to \infty$.
On the other hand, $\cal M$ appears to increase with $V$, as shown in Fig.~\ref{fig2_2}(b).
We note that in finite systems there is a clear separation in energy scale between zero modes and other eigenenergies, as shown in Figs.~\ref{fig2_2}(c) and~\ref{fig2_2}(d).
In particular, while the latter are roughly bounded by $|E_\alpha| \gtrsim V^{-1} \approx 10^{-4}$, we detect the former by setting the condition $|E_\alpha| < 10^{-14}$.

The results in Fig.~\ref{fig2_2} also show certain differences between the billiards B1 and B2.
An important observation is that in the billiard B1, the sublattice imbalance $\delta N$ is not only nonzero, but it appears to systematically increase with $V$.
In fact, we observe an exact relation ${\cal M} = \delta N$, which is shown by the agreement between open and filled symbols (circles and triangles) in Fig.~\ref{fig2_2}(b).
In general, the number of zero modes $\cal M$ (and hence $\delta N$) is much larger in the billiard B1 than in B2, and Fig.~\ref{fig2_2}(a) suggests that the fraction ${\cal M}/V$ in B1 (and hence $\delta N/V$) is well described by a power-law fit $\propto V^{-\zeta}$ with an exponent $\zeta\approx 0.5$.

As a consequence of the relation ${\cal M} = \delta N$ in the billiard B1, its number of zero modes is exactly given by the sublattice imbalance and hence $m=0$ in Eq.~(\ref{def_M_exact}).
The origin of the nontrivial dependence of $\delta N$ on $V$ in the billiard B1 ($\delta N \approx V^{1/2}$) is geometric: it is the boundary cut along the diagonal that yields an excess of the number of sites in one sublattice relative to the other sublattice.
This, in turn, unveils a generic mechanism for the emergence of zero modes in tight-binding billiards: if the boundaries of the lattice are shaped such that the number of sites in one sublattice is systematically larger than in another sublattice, hence $\delta N \gg 1$, then, as per Eq.~(\ref{def_M_exact}), it gives rise to a massive number of zero modes $\cal M$.

We stress, however, that a large sublattice imbalance $\delta N \gg 1$ may not be a necessary requirement for the emergence of a large number of zero modes.
In Fig.~\ref{fig2_2}(b) we show that in the billiard B2 the sublattice imbalance fluctuates around zero, $\delta N \approx 0$, while the number of zero modes $\cal M$ is much larger than $\delta N$ and it appears to increase with $V$, at least for the system sizes under investigation.
Understanding the origin of these differences between the billiards B1 and B2 is an open question for future work.

In passing, we note that recent work~\cite{schecter_iadecola_18} identified sufficient conditions for the emergence of a large number of zero modes in a certain class of quantum {\it many-body} Hamiltonians in a lattice: they require existence of an operator $\hat \Gamma'$ that anticommutes with the Hamiltonian, and a lattice inversion symmetry that commutes with $\hat \Gamma'$.
While the explicit expressions for the corresponding operators in quantum many-body Hamiltonians are different from those in the tight-binding billiards studied here, the outcomes of their action (i.e., emergence of zero modes) appear to be very similar.

One can also make further statements about the structure of the wavefunctions of zero modes.
Let us assume that one finds a zero mode $|\alpha\rangle$ that is of the form given by Eq.~(\ref{def_alpha}), i.e., it is a superposition of particle occupations on both sublattices.
As a consequence of Eq.~(\ref{def_H_Gamma_alpha}), if $|\alpha\rangle$ is a zero mode, the same holds true for $\hat\Gamma |\alpha\rangle$. 
One can then construct the symmetrized zero modes as
\begin{equation} \label{def_symmetrized_zeromodes}
    |\alpha_A\rangle \propto \left(|\alpha\rangle + \hat \Gamma |\alpha\rangle \right)\;,\;\;\;
    |\alpha_B\rangle \propto \left(|\alpha\rangle - \hat \Gamma |\alpha\rangle\right)\;,
\end{equation}
where $|\alpha_A\rangle$ only includes occupations in sublattice A and $|\alpha_B\rangle$ only in sublattice B.
These states are also eigenstates of the chiral operator $\hat\Gamma$ with eigenvalues $\gamma_A = 1$ and $\gamma_B = -1$,
\begin{equation} \label{def_eigenstates_gamma}
    \hat\Gamma|\alpha_A\rangle = \gamma_A|\alpha_A\rangle = |\alpha_A\rangle \;,\;\;\;
    \hat\Gamma|\alpha_B\rangle = \gamma_B|\alpha_B\rangle = -|\alpha_B\rangle \;.
\end{equation}
We hence refer to the single-particle states $|\alpha_A\rangle$ and $|\alpha_B\rangle$ at zero energy as chiral particles.
Note that at nonzero energy, $|\alpha_A\rangle$ and $|\alpha_B\rangle$ are not eigenstates of $\hat H$.

In Appendix~\ref{app:zeromodes} we show that for a given number of zero modes $\cal M$ as given by Eq.~(\ref{def_M_exact}), there are $\delta N + m$ zero modes that are confined into the sublattice A and $m$ zero modes confined into the sublattice B (see also~\cite{wilming_osborne_22}).
In terms of the eigenvalues $\gamma_A$, $\gamma_B$ of the chiral operator $\hat \Gamma$, see Eq.~(\ref{def_eigenstates_gamma}), one hence obtains the sum rule for all eigenvalues of the zero modes as
\begin{equation}
    \sum_{i=1}^{\delta N + m} \gamma_{A,i} + \sum_{j=1}^m \gamma_{B,j} = \delta N \;.
\end{equation}
Figures~\ref{fig14} and~\ref{fig13} of Appendix~\ref{app:zeromodes} show examples of the wavefunction amplitudes of some zero modes, and contrast them to wavefunction amplitudes of Hamiltonian eigenstates at nonzero energies.

\section{Single-particle energy spectrum} \label{sec:spectrum}

We now turn our attention to the properties of the single-particle energy spectrum of tight-binding billiards.
Some of its properties share similarities with spectra of free fermions on square lattices with regular boundaries such as those formed by a box.
For example, the spectrum is bounded to the interval $E_\alpha \in \left[-4, 4\right]$ and the density of states (not shown here) is also very similar in both cases.

\begin{figure}[!t]
\centering
\includegraphics[width=0.98\columnwidth]{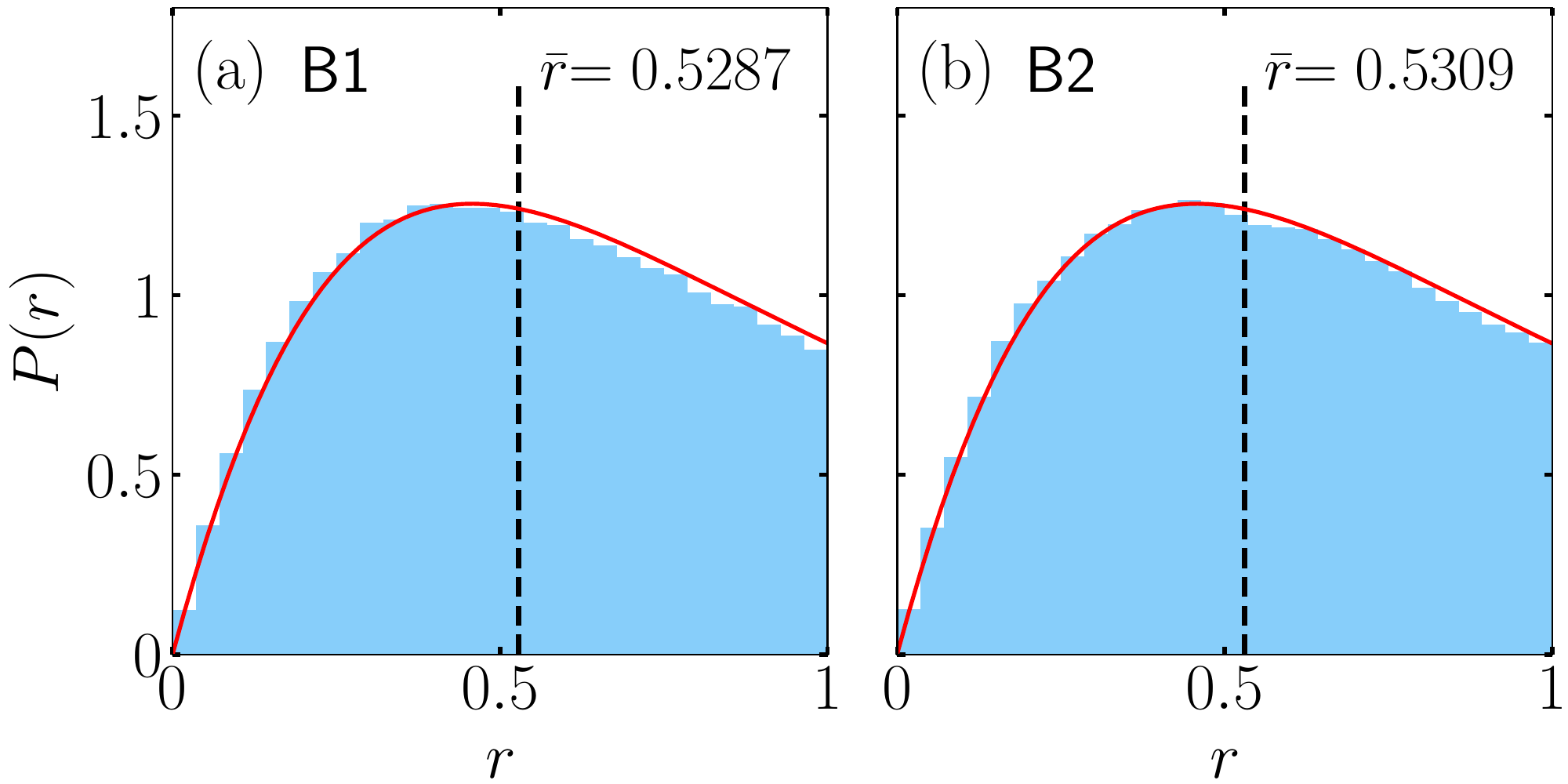}
\caption{
Probability density functions $P(r)$ of level spacing ratios in the billiards (a) B1 and (b) B2.
Results are obtained from $80\%$ of energy levels in the middle of the spectrum (excluding zero modes) and further averaged over 20 system sizes for $V \in [25000,30000]$.
Red solid line is the result from Eq.~(\ref{eq:rRMT}).
Vertical dashed lines are the mean values of $r$, which yield (a) $\bar r = 0.5287$ and (b) $\bar r = 0.5309$.
}\label{fig2}
\end{figure}

Here we study statistical properties of the energy spectrum of tight-binding billiards that may be fundamentally different from those in systems with regular boundaries.
We focus on the nearest level spacings $\delta_\alpha = E_\alpha - E_{\alpha-1}$, and we consider the ratio $r_\alpha = \min(\delta_{\alpha+1}, \delta_\alpha)/\max(\delta_{\alpha+1}, \delta_\alpha)$~\cite{oganesyan_huse_07}.
While $r_\alpha$ denotes the ratio for a particular target energy level $\alpha$, we denote $\bar r$ as the ratio after the averaging over different levels within the same system and after the averaging over different system sizes.

We first consider the probability density function $P(r)$ of $80\%$ of levels around the middle of the spectrum.
We exclude information from zero modes, for which the ratio is not well defined, and average the results over $20$ system sizes spanning from approximately $V = 25000$ to $V = 30000$ (the average volume is 27829.45 in the billiard B1 at $R=L/2$ and 27485.95 in the billiard B2).
The purpose of the latter averaging is to minimize discretization effects when designing tight-binding billiards from continuous billiards, as described in Sec.~\ref{sec:models}.
Results in both billiards are shown as histograms in Fig.~\ref{fig2}.
The red solid line is an exact prediction within a Gaussian orthogonal ensemble (GOE)  for $3 \times 3$ matrices~\cite{atas_bogomolny_13},
\begin{equation}
\label{eq:rRMT}
P(r)=\frac{27}{4} \frac{r+r^{2}}{\left(1+r+r^{2}\right)^{5 / 2}} \;.
\end{equation}
The agreement between the numerical results and the prediction from Eq.~(\ref{eq:rRMT}) is very good.
Moreover, the average value that we obtain in the billiard B2 in Fig.~\ref{fig2}(b) is $\bar r = 0.5309$, which is remarkably close to the numerical predictions within the GOE for asymptotically large matrices, $\bar{r}_{\mathrm{GOE}} = 0.5307$~\cite{atas_bogomolny_13}.
In the billiard B1 in Fig.~\ref{fig2}(a) we get $\bar r = 0.5287$, which is also close to $\bar{r}_{\mathrm{GOE}}$, but still not as close as the results in the billiard B2.

\begin{figure}[!t]
\centering
\includegraphics[width=0.98\columnwidth]{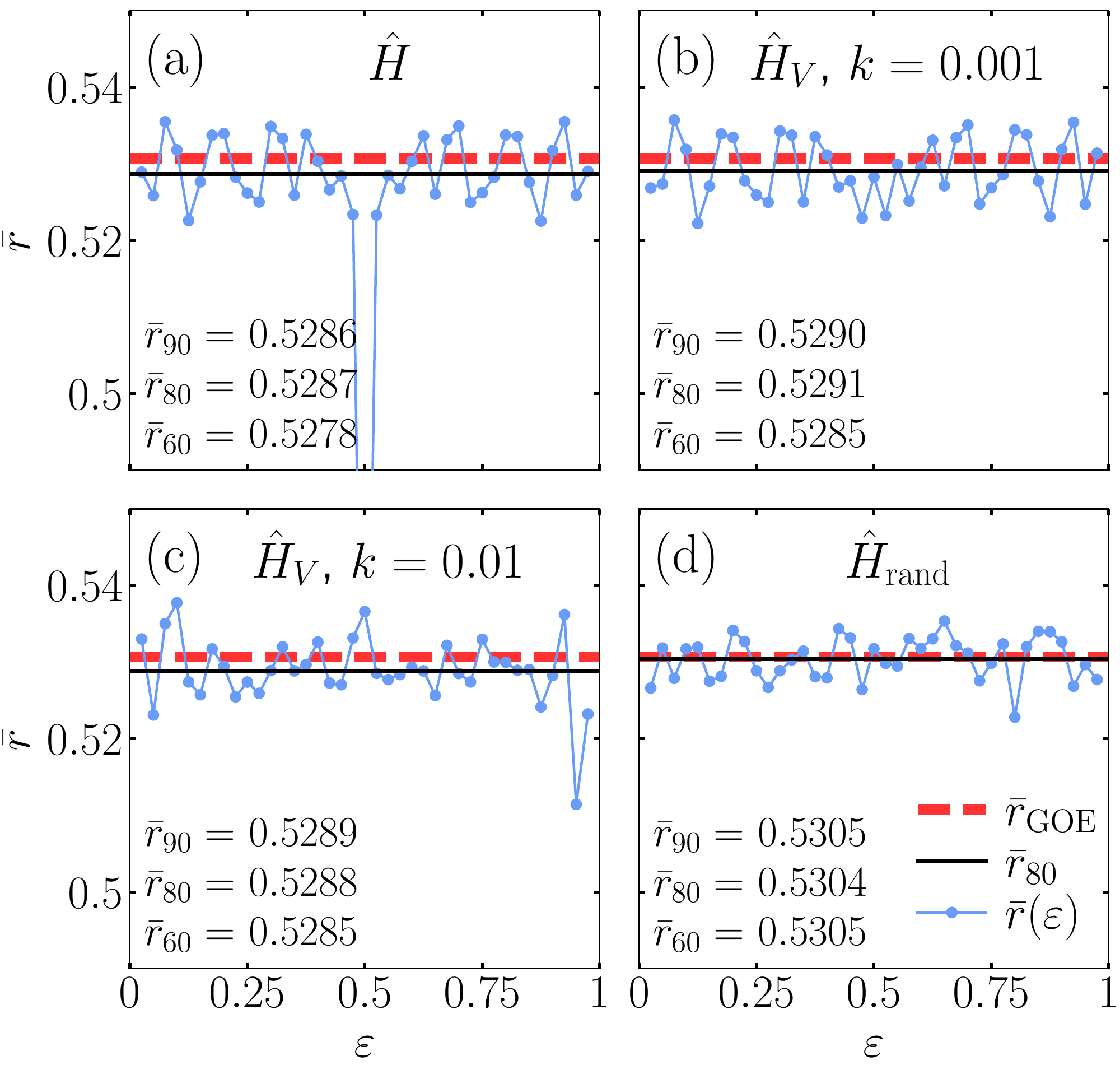}
\caption{
Level spacing $\bar r$ versus the normalized energy $\varepsilon$ in the billiard B1.
Points connected by blue lines are the values of $\bar r$ at the given $\varepsilon$, averaged over 500 neighboring energies.
Red dashed line is the GOE prediction $\bar r_{\rm GOE}=0.5307$~\cite{atas_bogomolny_13}.
In the legends we list the average values $\bar r_{90}$, $\bar r_{80}$ and $\bar r_{60}$, which, respectively, correspond to $90\%$, $80\%$ and $60\%$ of all levels around the middle of the spectrum, excluding zero modes.
The black solid line represents $\bar{r}_{80}$.
Results are shown for the following Hamiltonians:
(a) $\hat H$ from~(\ref{eq:billiard}),
(b) $\hat H_k$ from~(\ref{eq:billiard_ho}) at $k = 0.001$,
(c) $\hat H_k$ from~(\ref{eq:billiard_ho}) at $k = 0.01$,
(d) $\hat H_{\rm rand}$ from~(\ref{eq:billiard_random}) at $W=0.5$.
Numerical results in (a)-(c) are averaged over 20 system sizes within $V=25000$ and $30000$, while in (d) the averaging is performed over $20$ different disorder realizations at the fixed system size $V = 27918$.
}\label{fig2b}
\end{figure}

\begin{figure}[!t]
\centering
\includegraphics[width=0.98\columnwidth]{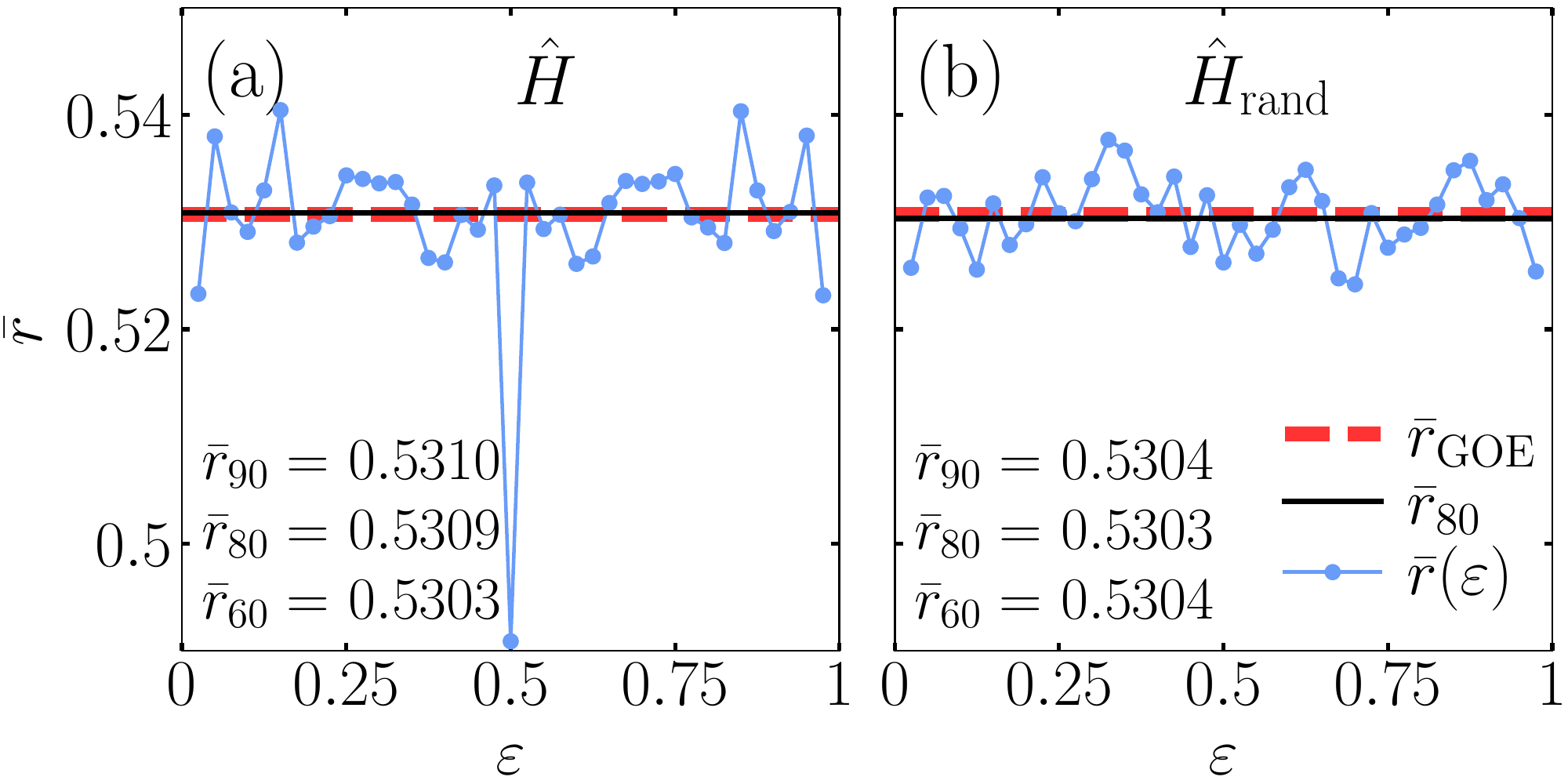}
\caption{
Level spacing $\bar r$ versus the normalized energy $\varepsilon$ in the billiard B2.
Results are analogous to those in Fig.~\ref{fig2b}, and shown for 
(a) $\hat H$ from~(\ref{eq:billiard}), and
(b) $\hat H_{\rm rand}$ from~(\ref{eq:billiard_random}) at $W=0.5$, for $V = 27600$.
}\label{fig2c}
\end{figure}

Next we focus on the energy dependence of the level spacing ratio $r$, plotted as a function of normalized energy $\varepsilon = (E-E_{\rm min})/(E_{\mathrm{max}} - E_{\mathrm{min}})$.
Results are shown as points connected by blue lines in Figs.~\ref{fig2b} and~\ref{fig2c}.
We plot 39 points and each point is an average over 500 neighboring energies.
We set $r=0$ for zero modes that belong to the point $\varepsilon = 0.5$ in Figs.~\ref{fig2b}(a) and~\ref{fig2c}(a), while other points do not contain contributions from zero modes.
These results are compared to two energy independent values:
$\bar r_{\rm GOE}$ (red dashed lines) and $\bar r_{80}$ (black solid lines), which is the average over $80\%$ of all levels around the middle of the spectrum, excluding contributions from zero modes.
All numerical results, unless stated otherwise, are also averaged over 20 system sizes within $V=25000$ and $V=30000$ as in Fig.~\ref{fig2}.

Results in the billiard B1 and the Hamiltonian $\hat H$ from Eq.~(\ref{eq:billiard}), shown in Fig.~\ref{fig2b}(a), exhibit a reasonably good agreement with the GOE prediction $\bar r_{\rm GOE}$ in a wide interval of normalized energies $\varepsilon$.
However, the average values of $r$ exhibit a small difference compared to $\bar r_{\rm GOE}$ that appears to be insensitive to the fraction of levels included in the average.
In all panels of Figs.~\ref{fig2b} and~\ref{fig2c} we provide the averages $\bar r_{90}$, $\bar r_{80}$ and $\bar r_{60}$, which, respectively, correspond to $90\%$, $80\%$ and $60\%$ of all levels around the middle of the spectrum, excluding zero modes.

We complement these results by performing an identical analysis of the Hamiltonians $\hat H_k$ and $\hat H_{\rm rand}$ from Eqs.~(\ref{eq:billiard_ho}) and~(\ref{eq:billiard_random}), respectively, which do not exhibit any degeneracy and reflection symmetry of the energy spectrum.
As a consequence, $\bar r$ ceases to be symmetric around $\varepsilon = 0.5$ [as it is in Fig.~\ref{fig2b}(a)].
Results for $\hat H_k$ at $k=0.001$ and $k=0.01$ are shown in Figs.~\ref{fig2b}(b) and~\ref{fig2b}(c), respectively.
The chosen values of harmonic potentials $k$ are small such that their main contribution is to remove degeneracies and reflection symmetry of the spectrum.
Nevertheless, we do not observe any significant quantitative impact of this contribution to the averaged level spacing ratios, and results at larger $k$ (not shown) exhibit even larger deviation from the GOE predictions.
On the other hand, adding a weak random disorder $W=0.5$, see Fig.~\ref{fig2b}(d), brings the averages closer to the GOE prediction $\bar r_{\rm GOE}$.
In fact, they almost perfectly agree with $\bar r_{\rm GOE}$ as the agreement is on the fourth digit.
We note that the agreement with the GOE predictions for the 2D Anderson model is likely an effect of the localization length being much larger than the lattice size, since in the thermodynamic limit the system is expected to be localized for any nonzero $W$~\cite{abrahams_anderson_79}.

Results in the billiard B2 in Fig.~\ref{fig2c} appear to be consistent with the results in the billiard B1 in Fig.~\ref{fig2b}, at least in the sense that the energy dependence of $\bar r$ is rather insignificant in both cases, and the fluctuations about the average are comparable.
However, the quantitative agreement of the averages with the GOE prediction is clearly better in the billiard B2.
Figure~\ref{fig2c}(a) shows that the agreement is nearly perfect (on four digits) for the Hamiltonian $\hat H$ that does not contain any on-site potentials, and the high level of agreement persists upon addition of a weak disorder, see Fig.~\ref{fig2c}(b).

These results suggest that the statistical properties of an overwhelming majority of energy levels complies with the GOE predictions.
They also offer two further insights.
First, there exist a small but persistent difference between the average level spacing ratios of both billiards, since, at least for the system sizes under considerations, the results in billiard B2 are always closer to the GOE prediction.
Second, the presence of degeneracies (zero modes) and reflection symmetry in the energy spectrum does not appear to crucially impact the degree of agreement (of non-degenerate levels) with the GOE predictions.

\section{Bipartite entanglement entropies of many-body eigenstates} \label{sec:entanglement}

We next study the structure of Hamiltonian eigenstates.
In contrast to other sections of this work that are devoted to single-particle properties, here we study properties of many-body eigenstates.
In particular, we study the entanglement content of these states as measured by the bipartite entanglement entropies.

\begin{figure}[!t]
\centering
\includegraphics[width=0.98\columnwidth]{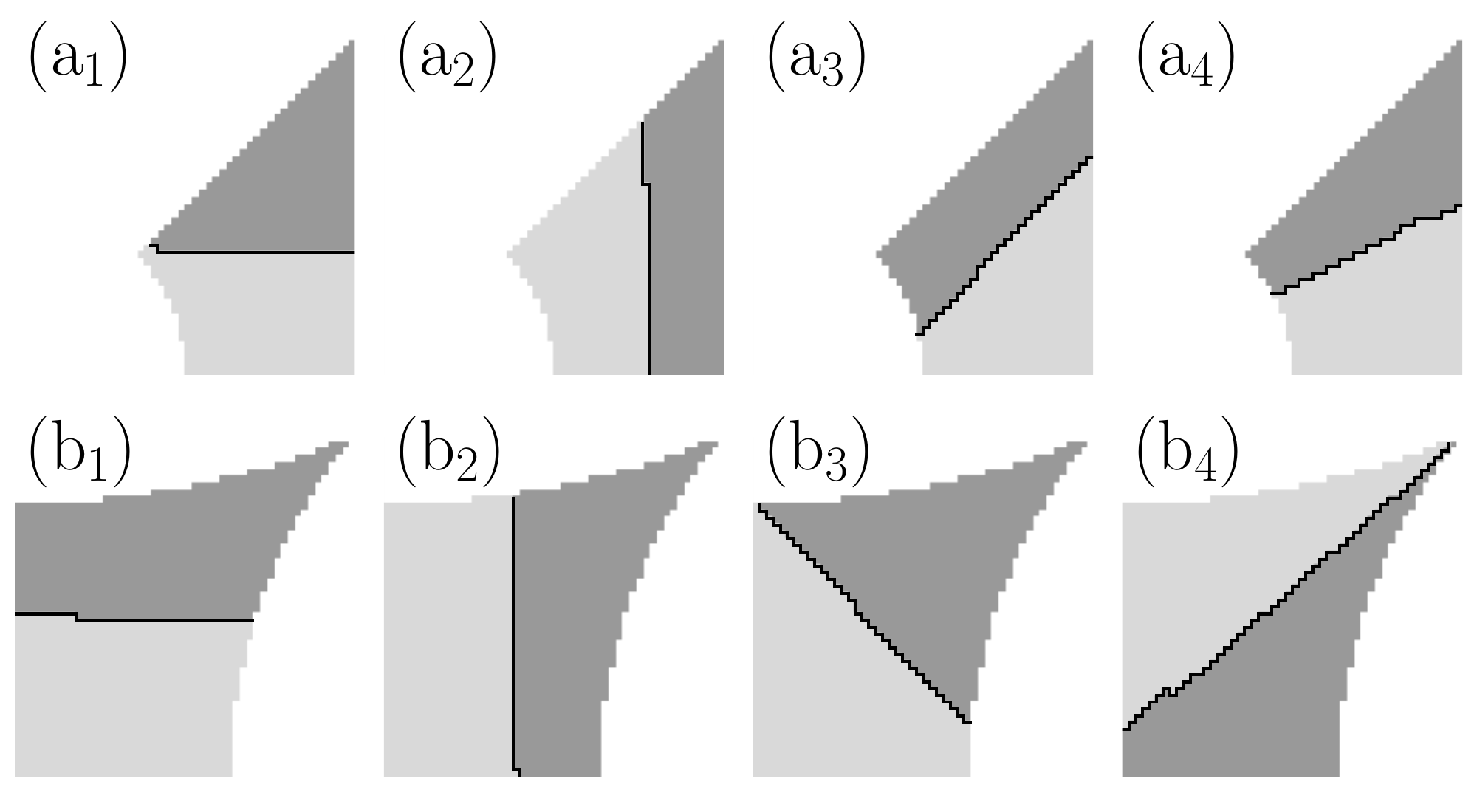}
\caption{
Lattice bipartitions applied in the study of entanglement entropies in Sec.~\ref{sec:entanglement}.
Upper row (a$_1$)-(a$_4$): four different bipartitions of the billiard B1.
Lower row (b$_1$)-(b$_4$): four different bipartitions of the billiard B2.
Protocols for designing these bipartitions are explained in Appendix~\ref{app:bipartitions}.
}\label{fig3}
\end{figure}

We define the von Neumann entanglement entropy of a many-body eigenstate $|m\rangle$ as
\begin{equation} \label{def_Sm}
    S_m = - {\rm Tr} \{ \hat \rho_{\cal A} \ln \hat\rho_{\cal A} \}\,,
\end{equation}
where $\hat\rho_{\cal A} = {\rm Tr}_{\cal B}\{\hat \rho_m\}$ is the reduced density matrix of $|m\rangle$ in a subsystem $\cal A$, and $\hat \rho_m = |m\rangle\langle m|$.
The 2nd R\' enyi entanglement entropy of a many-body eigenstate $|m\rangle$ is then defined as
\begin{equation} \label{def_Sm2}
    S_m^{(2)} = - \ln[{\rm Tr}\{ \hat \rho_{\cal A}^2\}]\,.
\end{equation}
In the tight-binding billiards considered here, there is no obvious preferential way how to chose a lattice bipartition.
In what follows we focus on bipartitions for which the number $V_{\cal A}$ of lattice sites in subsystem ${\cal A}$ is identical to the number $V_{\cal B} = V-V_{\cal A}$ of lattice sites in subsystem ${\cal B}$.
[If $V$ is odd, we set $V_{\cal A}$ as the integer part of $V/2$.]
In each tight-binding billiard we consider four bipartitions that are sketched in Fig.~\ref{fig3}.
A proper scaling analysis of entanglement entropies requires definitions of bipartitions that can be applied to an arbitrary large system, and we explain particular protocols to achieve this goal in Appendix~\ref{app:bipartitions}.
We note that the lattice bipartitions used in the context of entanglement studies in this section are unrelated to the discussion of sublattices with checkerboard patterns in Sec.~\ref{sec:general} and Fig.~\ref{fig_sublattice}(a), even though entanglement studies of the latter bipartitions may also represent an interesting subject~\cite{wilming_osborne_22}.

The general motivation for studying the eigenstate entanglement entropies stems from a recent conjecture~\cite{leblond_mallayya_19} (see also~\cite{vidmar_hackl_17, vidmar_hackl_18, lydzba_rigol_20, bianchi_hackl_22}) that the entanglement entropies of typical excited eigenstates represent an efficient tool to distinguish quantum-chaotic interacting Hamiltonians from quadratic and integrable interacting Hamiltonians.
A particular motivation for studies of tight-binding billiards is whether the volume-law contribution to the entanglement entropy is quantitatively described by the analytical expressions that we introduce in Eqs.~(\ref{def_S_rmt}) and~(\ref{def_S2_rmt}), which were conjectured to describe the results in quantum-chaotic quadratic Hamiltonians.
Recently, several works have started to explore entanglement properties of excited eigenstates of quadratic and integrable interacting Hamiltonians~\cite{alba_fagotti_09, moelter_barthel_14, storms_singh_14, lai_yang_15, nandy_sen_16, vidmar_hackl_17, vidmar_hackl_18, liu_chen_18, hackl_vidmar_19, zhang_vidmar_18, jafarizadeh_rajabpour_19, Roy_2019, modak_nag_20, lydzba_rigol_20, lydzba_rigol_21, modak_mandal_21, bhattacharjee_nandy_21, bernard_piroli_21, miao_barthel_21, miao_barthel_22, zhang_rajabpour_22, kumari_alhambra_22, murciano_calabrese_22}.

We are interested in the average bipartite entanglement entropies over all many-body eigenstates.
However, the number of the latter increases exponentially as $\propto 2^V$, which poses serious numerical difficulties for systems of size $V \approx 40$ or larger.
We therefore approximate this average by an average over $M$ randomly selected many-body eigenstates, and we define
\begin{equation} \label{def_Saverage}
    \overline{S} = \frac{1}{M} \sum_{m=1}^M S_m \;, \;\;\; 
    \overline{S^{(2)}} = \frac{1}{M} \sum_{m=1}^M S_m^{(2)} \;.
\end{equation}
We choose $M= 1000$ in our studies, and allow for both non-zero energy single-particle eigenstates as well as the zero modes to form a many-body eigenstate.
In addition to the average over randomly selected eigenstates within the same system, as given by Eq.~(\ref{def_Saverage}), we also average the results over 5 systems with similar numbers of lattice sites, using the same labels $\overline{S}$ and $\overline{S^{(2)}}$.
This procedure is analogous to the one performed for the average level spacing ratio in Sec.~\ref{sec:spectrum}, and it is used to smoothen fluctuations that emerge due to discretization effects in the construction of tight-binding billiards.
Throughout the work we refer to the averages introduced above as the average eigenstate entanglement entropies.

We note that in the actual numerical calculations of $S_m$ and $S_m^{(2)}$ one does not need to calculate the reduced density matrix of a many-body eigenstate $|m\rangle$, as indicated in Eqs.~(\ref{def_Sm}) and~(\ref{def_Sm2}).
For quadratic Hamiltonians as considered in this work, it suffices to calculate the so-called one-body correlation matrix, which includes matrix elements of one-body observables only~\cite{peschel_03, peschel_eisler_09}.
This procedure for calculating the bipartite entanglement entropies is well established (see, e.g., Refs.~\cite{lydzba_rigol_20, lydzba_rigol_21, bianchi_hackl_22}), and we summarize it for convenience in Appendix~\ref{app:entropies}.

We compare results to the analytical predictions for the volume-law contribution to the entanglement entropies.
For the von Neumann entanglement entropy, the prediction is
\begin{equation} \label{def_S_rmt}
    \overline{\cal S} = \left(1-\frac{1+f^{-1}\left(1-f\right) \ln\left(1-f\right)}{\ln 2}\right)V_{\cal A} \ln 2 \,,
\end{equation}
where $f=V_{\cal A}/V$ is the fraction of the subsystem volume relative to the total volume. [Eq.~(\ref{def_S_rmt}) is valid at $f \leq 1/2$, while for $1/2<f\leq 1$ one needs to replace $V_{\cal A} \to V - V_{\cal A}$ and $f \to (1-f)$.]
This result was first derived in Ref.~\cite{lydzba_rigol_20} using the assumption that the coefficients of single-particle eigenstates are normally distributed random numbers.
An identical result was obtained in an ensemble of pure fermionic Gaussian states~\cite{bianchi_hackl_21, bianchi_hackl_22}.
For the 2nd R\' enyi entanglement entropy there is to our knowledge no simple closed-form expression $\overline{{\cal S}^{(2)}}(f)$, apart from the $f=1/2$ point~\cite{zhang_liu_20, lydzba_rigol_21}.
Recently, by using the distribution of eigenvalues of the restricted one-body correlation matrix~\cite{liu_chen_18}, $\overline{{\cal S}^{(2)}}(f)$ was expressed as an infinite series of hypergeometric functions, see Eq.~(28) in~\cite{lydzba_rigol_21}, and as part of a general expression for the $n$-th R\' enyi ground-state entanglement entropy in the SYK2 model, see Eq.~(41) in~\cite{zhang_liu_20}.
Here we argue that both expressions are equivalent and we use the result from Ref.~\cite{zhang_liu_20} to obtain a closed-form expression.
The latter is
\begin{widetext}
\begin{equation} \label{def_S2_rmt}
    \overline{{\cal S}^{(2)}} = \left( 1- \frac{\log_2 \left(\frac{\sqrt{1+4f(1-f)}+1}{2}\right) - (1-2f) \log_2 \left(1+ \frac{\sqrt{1+4f(1-f)} - 1}{2(1-f)} \right) }
    {f} \right) V_{\cal A} \ln 2 \;,
\end{equation}
\end{widetext}
and its derivation is carried out in Appendix~\ref{app:renyi}.
Moreover, it was shown in~\cite{lydzba_rigol_21} that $\overline{{\cal S}^{(2)}}$ accurately describes the volume-law contribution to the average eigenstate entanglement entropy in quantum-chaotic quadratic Hamiltonians such as the 3D Anderson model and the SYK2 models.
We hence use the result from Eq.~(\ref{def_S2_rmt}) as a reference point for the average eigenstate entanglement entropy in tight-binding billiards.

\subsection{Scaling of the volume-law coefficients}

\begin{figure}[!t]
\centering
\includegraphics[width=0.98\columnwidth]{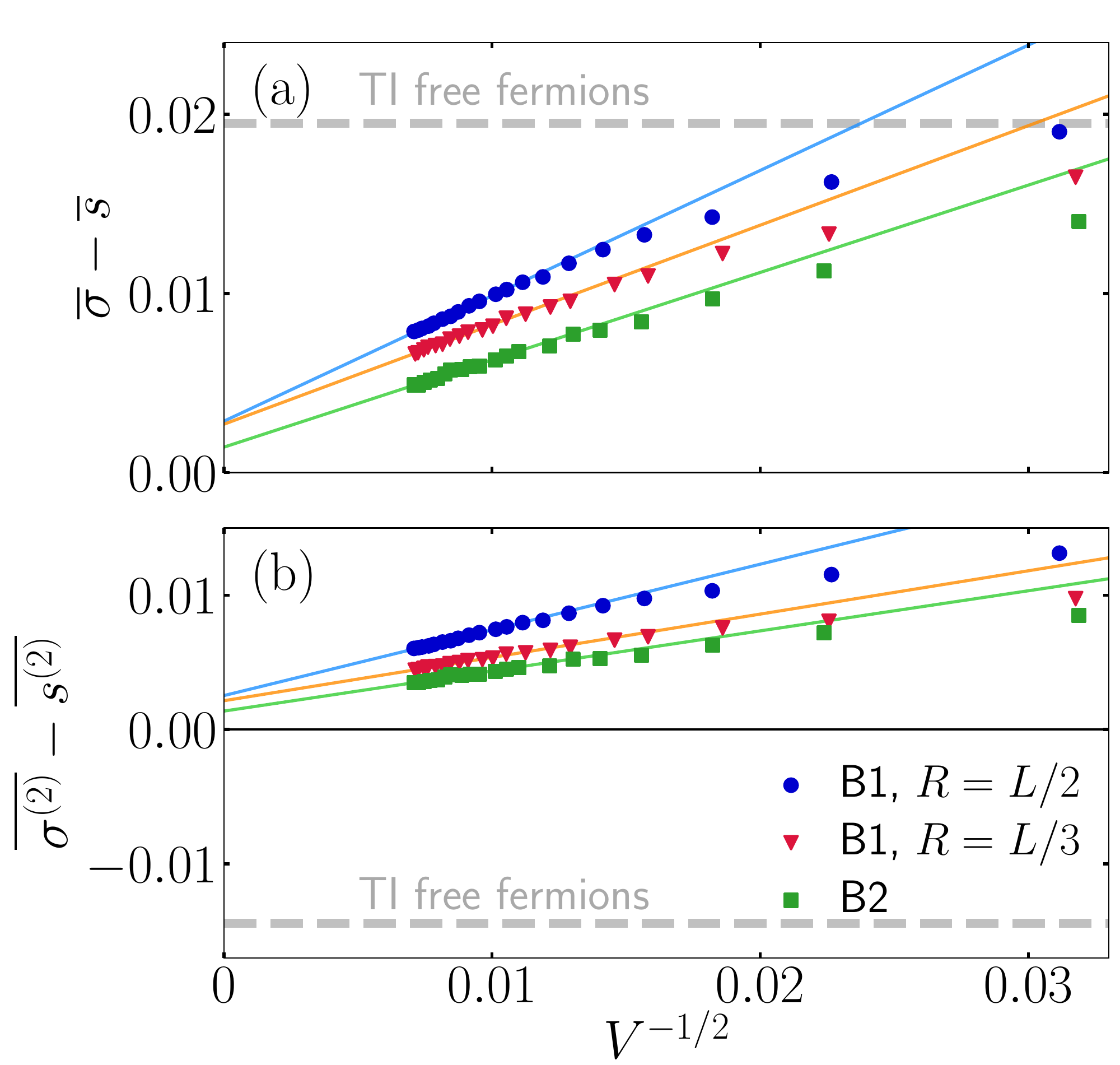}
\caption{
Differences of the volume-law coefficients, see Eqs.~(\ref{def_volume_coef}) and~(\ref{def_volume_coefficient_rmt}), versus $V^{-1/2}$, in the billiards B1 at $R=L/2$ and $R=L/3$, and B2.
The billiard bipartitions from Fig.~\ref{fig3} are ${\rm a}_1$ in B1 and ${\rm b}_1$ in B2.
Results are shown (a) for the von Neumann entropy, $\overline{\sigma} - \overline{s}$, and (b) for the 2nd R\' enyi entropy, $\overline{\sigma^{(2)}} - \overline{s^{(2)}}$.
Each symbol represents the average over 5 successive system sizes, as explained in the main text, and the volume $V$ is then the corresponding average.
Solid lines are two parameter fits of $g(V)$ from Eq.~(\ref{def_g_fit}) at $\eta=2$ to the results at $V \geq 11000$.
The value of the off-set parameter $a_0$ in $g(V)$ is in all cases $a_0 < 0.003$.
Horizontal dashed lines represent the differences between the volume-law coefficients $\overline{\sigma}=0.5573$ and $\overline{\sigma^{(2)}} = 0.4569$ from Eq.~(\ref{def_volume_coefficient_rmt}) at $f=1/2$, and the corresponding results for translationally-invariant (TI) free fermions in lattices with regular boundaries,
$\overline{s}_{\rm TI} = 0.5378$~\cite{vidmar_hackl_17} and $\overline{s^{(2)}}_{\rm TI} = 0.4713$~\cite{lydzba_rigol_21}, see also Appendix~\ref{app:Sfree}.
}\label{fig4}
\end{figure}

\begin{figure}[!t]
\centering
\includegraphics[width=0.98\columnwidth]{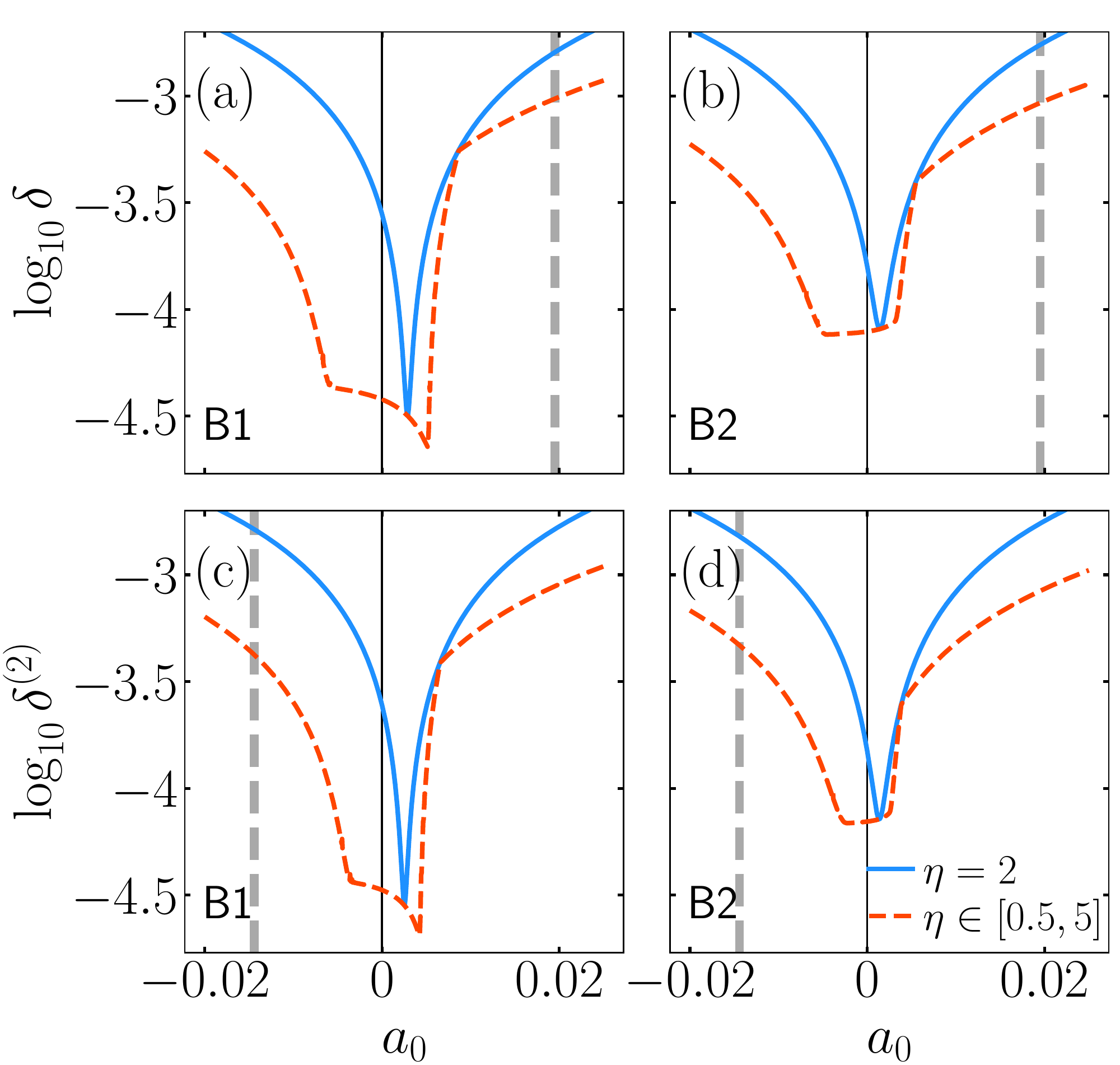}
\caption{
Quality of the fits to the numerical results (symbols) in Fig.~\ref{fig4} in the billiards B1 at $R=L/2$, and B2.
We plot $\log_{10}\delta$ of the von Neumann entropy, see Eq.~(\ref{def_delta_vN}), in the billiards (a) B1 and (b) B2, and $\log_{10}\delta^{(2)}$ of the 2nd R\' enyi entropy, see Eq.~(\ref{def_delta_Renyi}), in the billiards (c) B1 and (d) B2.
We first fix the parameter $a_0$ in the fitting function $g(V)$ in Eq.~(\ref{def_g_fit}) and then calculate $\delta$ and $\delta^{(2)}$ using results for the best fit of the parameter $a_1$ at $V \geq 11000$.
Solid lines are obtained by fixing $\eta=2$ in Eq.~(\ref{def_g_fit}), while the thin dashed lines correspond to considering $\eta$ as a fitting parameter within the interval $\eta \in [0.5,5]$.
Vertical dashed lines are identical to the horizontal dashed lines in Fig.~\ref{fig4}.
They represent the differences between analytical results from Eq.~(\ref{def_volume_coefficient_rmt}) at $f=1/2$, and the corresponding results for TI free fermions in lattices with regular boundaries.
}\label{fig5}
\end{figure}

In Figs.~\ref{fig4} and~\ref{fig5} we compare the numerical results from Eq.~(\ref{def_Saverage}) to the predictions from Eqs.~(\ref{def_S_rmt}) and~(\ref{def_S2_rmt}).
Since our main interest is in the leading contribution to the entanglement entropy, which scales with the volume $V_{\cal A}$ of the subsystem $\cal A$, we define the average entanglement entropy density (i.e., the volume-law coefficient) as
\begin{equation} \label{def_volume_coef}
    \overline{s} = \frac{\overline{S}}{V_{\cal A} \ln 2}\;,\;\;\;
    \overline{s^{(2)}} = \frac{\overline{S^{(2)}}}{V_{\cal A} \ln 2} \;
\end{equation}
for the entanglement entropies from Eq.~(\ref{def_Saverage}), and 
\begin{equation} \label{def_volume_coefficient_rmt}
    \overline{\sigma} = \frac{\overline{\cal S}}{V_{\cal A} \ln 2}\;,\;\;\;
    \overline{\sigma^{(2)}} = \frac{\overline{{\cal S}^{(2)}}}{V_{\cal A} \ln 2} \;
\end{equation}
for the entanglement entropies from Eqs.~(\ref{def_S_rmt}) and~(\ref{def_S2_rmt}), respectively.
The comparison in Figs.~\ref{fig4} and~\ref{fig5} is carried out for the bipartitions (a$_1$) and (b$_1$) from Fig.~\ref{fig3}, while other bipartitions are studied in Sec.~\ref{sec:bipartitions}.
This choice of bipartitions correspond to $f=1/2$ in Eqs.~(\ref{def_S_rmt}) and~(\ref{def_S2_rmt}) if $V$ is even and $f=1/2-1/(2V)$ if $V$ is odd.
At $f=1/2$, the result is $\overline{\sigma} = 0.5573$ and $\overline{\sigma^{(2)}} = 0.4569$.

Figures~\ref{fig4}(a) and~\ref{fig4}(b) show the scaling of the differences of volume-law coefficients $\overline{\sigma}-\overline{s}$ and $\overline{\sigma^{(2)}}-\overline{s^{(2)}}$, respectively, with the inverse volume.
We observe that the differences are well described by a function
\begin{equation} \label{def_g_fit}
g(V) = a_0 + a_1 V^{-1/\eta} \;,
\end{equation}
see the lines in Fig.~\ref{fig4}, where $a_0$ and $a_1$ are fitting parameters and we fix $\eta=2$.
The most important result of the fitting analysis is that the offset $a_0$ is very small for all cases under consideration.
In particular, the values of $a_0$ are of the order $10^{-3}$, which is a strong indication of the relevance of predictions in Eqs.~(\ref{def_S_rmt}) and~(\ref{def_S2_rmt}) for the average eigenstate entanglement entropies of our tight-binding billiards.
The scaling $\propto V^{-1/2}$ of the entanglement entropy density suggests that the dominant subleading term to the entanglement entropy scales with the linear dimension, $\propto V^{1/2} \approx L$.
We note, however, that a similar scaling $\propto V^{-1/2}$ of the entanglement entropy density was also observed in 3D Anderson models~\cite{lydzba_rigol_21}, as well as in 1D systems of interacting hard-core bosons in the integrable regime~\cite{leblond_mallayya_19} and in the quantum-chaotic regime away from half-filling~\cite{vidmar_rigol_17}.

A more precise analysis of the differences between numerical results in tight-binding billiards and the predictions from Eqs.~(\ref{def_S_rmt}) and~(\ref{def_S2_rmt}) is carried out in Fig.~\ref{fig5}.
We define the function
\begin{equation} \label{def_delta_vN}
    \delta^2 = \frac{1}{N} \sum_{j=1}^N \left[\overline{\sigma} - \overline{s}(V_j) - g(V_j)\right]^2
\end{equation}
for the von Neumann entropy, and the function
\begin{equation} \label{def_delta_Renyi}
    (\delta^{(2)})^2 = \frac{1}{N} \sum_{j=1}^N \left[\overline{\sigma^{(2)}} - \overline{s^{(2)}}(V_j) - g(V_j)\right]^2
\end{equation}
for the 2nd R\' enyi entropy, where $N$ is the number of fitting points.
They both quantify the quality of the fitting function $g(V)$ from Eq.~(\ref{def_g_fit}).

Figure~\ref{fig5} shows the quality of the fits $\delta$ and $\delta^{(2)}$ for the numerical results in Fig.~\ref{fig4}.
Results are plotted as a function of the off-set $a_0$ in the fitting function $g(V)$ from Eq.~(\ref{def_g_fit}).
This implies that the optimal value of $a_0$ corresponds to the minimum of $\delta$ and $\delta^{(2)}$.
We observe $a_0 \approx 0$ in all cases under consideration, which suggests that the volume-law coefficient of the average eigenstate entanglement entropies is to high accuracy provided by the predictions from Eq.~(\ref{def_volume_coefficient_rmt}).
The accuracy is of the order of $10^{-3}$.

Recent studies have shown that the volume-law coefficients of the average eigenstate entanglement entropies of translationally-invariant (TI) free fermions in lattices with regular boundaries~\cite{vidmar_hackl_17, vidmar_hackl_18, hackl_vidmar_19, lydzba_rigol_20, lydzba_rigol_21} do not comply with predictions from Eqs.~(\ref{def_S_rmt}) and~(\ref{def_S2_rmt}).
Instead, their entanglement entropies $\overline{S}_{\rm TI}$ and $\overline{S^{(2)}}_{\rm TI}$ appear to follow another universal function of subsystem fraction $f$, whose closed-form analytical form is still unknown (see~\cite{hackl_vidmar_19} for some attempts in this direction).
For example, the volume-law coefficients at $f=1/2$ are $\overline{s}_{\rm TI} = 0.5378$~\cite{vidmar_hackl_17} and $\overline{s^{(2)}}_{\rm TI} = 0.4713$~\cite{lydzba_rigol_21}.
We note that while Refs.~\cite{vidmar_hackl_17, vidmar_hackl_18, hackl_vidmar_19, lydzba_rigol_20, lydzba_rigol_21} only considered TI free fermions in one- or three-dimensional lattices, we show in Appendix~\ref{app:Sfree} that their results are consistent with those for square lattices.

The deviations between the results for TI free fermions in lattices with regular boundaries and predictions from Eqs.~(\ref{def_S_rmt}) and~(\ref{def_S2_rmt}) are shown with dashed horizontal lines (vertical lines) in Fig.~\ref{fig4} (Fig.~\ref{fig5}).
In all cases under consideration, results for TI free fermions in lattices with regular boundaries do not appear to apply to the results for tight-binding billiards.

\subsection{Impact of bipartitions} \label{sec:bipartitions}

\begin{figure}[!t]
\centering
\includegraphics[width=0.98\columnwidth]{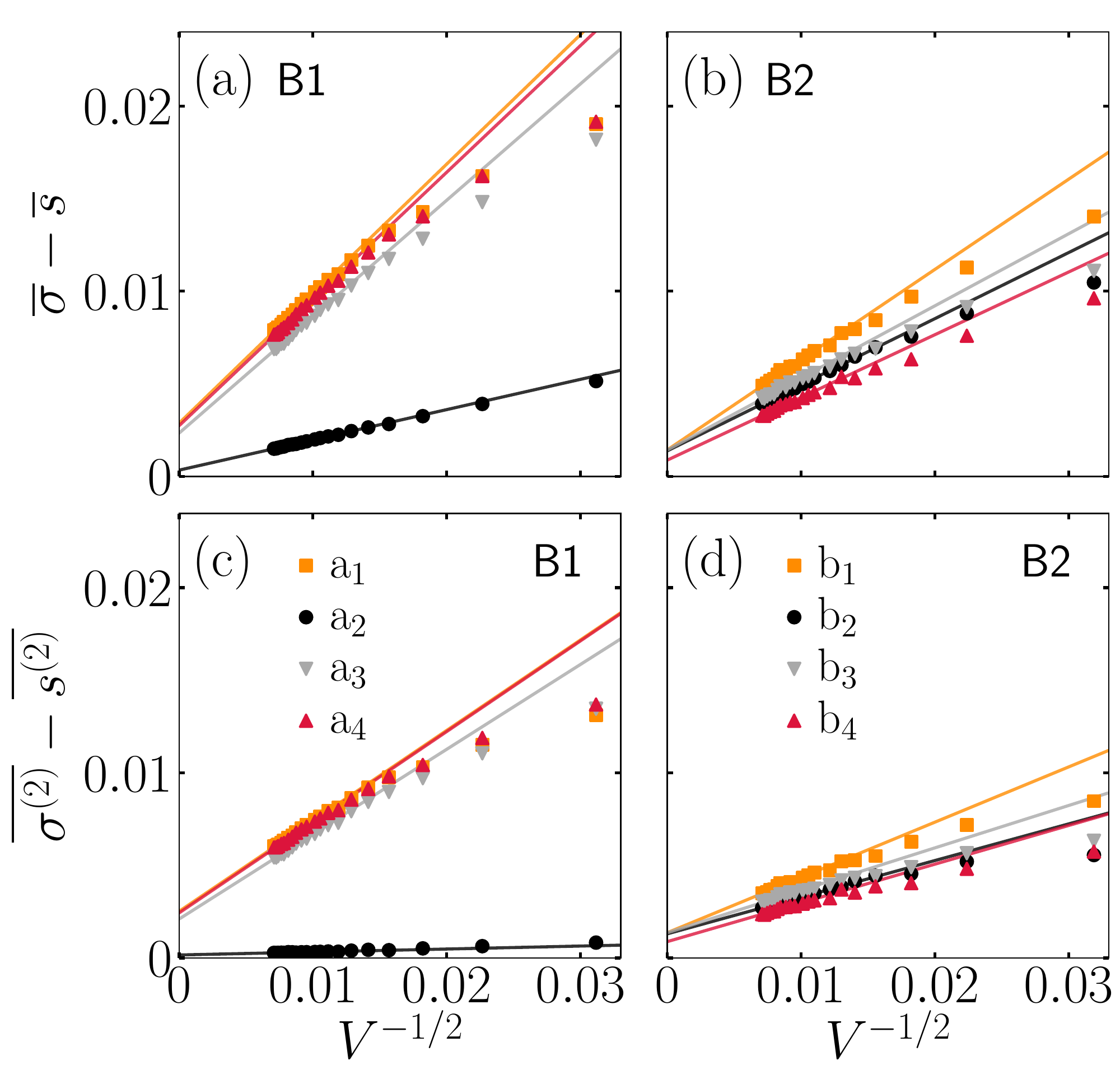}
\caption{
Differences of the volume-law coefficients, see Eqs.~(\ref{def_volume_coef}) and~(\ref{def_volume_coefficient_rmt}), versus $V^{-1/2}$, in the billiards (a, c) B1 and (b, d) B2, and different billiard bipartitions from Fig.~\ref{fig3}.
Results are shown (a, b) for the von Neumann entropy, $\overline{\sigma} - \overline{s}$, and (c, d) for the 2nd R\' enyi entropy, $\overline{\sigma^{(2)}} - \overline{s^{(2)}}$.
Each symbol represents the average over 5 successive system sizes, as explained in the main text, and the volume $V$ is then the corresponding average.
Solid lines are two parameter fits of $g(V)$ from Eq.~(\ref{def_g_fit}) at $\eta=2$ to the results at $V \geq 11000$.
The value of the off-set parameter $a_0$ in $g(V)$ is in all cases $a_0 < 3 \cdot 10^{-3}$.
In the billiard B1 and the bipartition $\mathrm{a}_2$ we get $a_0 \lesssim 3 \cdot 10^{-4}$
}\label{fig6}
\end{figure}

We complement previous results with those for other lattice bipartitions from Fig.~\ref{fig3}.
Figure~\ref{fig6} shows the finite-size scaling of the deviations of the volume-law coefficients~(\ref{def_volume_coef}) from the analytical predictions~(\ref{def_volume_coefficient_rmt}).
In general we observe that in all cases the volume-law coefficients do approach values that are close to the analytical predictions, with comparable accuracy as in the analysis in Fig.~\ref{fig5}.
We also make two additional observations.
The first is that the finite-size deviations are in principle smaller in the billiard B2.
This may be consistent with a smaller number of zero modes in this billiard, as discussed in the context of Fig.~\ref{fig2_2}.
The second is that there exist a bipartition in the billiard B1 (referred to as a$_2$ in Fig.~\ref{fig3}), for which the numerical results for the average eigenstate entanglement entropy almost precisely follow the analytical predictions from Eq.~(\ref{def_volume_coefficient_rmt}) already in rather small systems, see Figs.~\ref{fig6}(a) and~\ref{fig6}(c).
While we are not able to provide a detailed explanation of this observation, we also note that the difference of the sublattice imbalance $\delta N$~(\ref{def_delta_N}) between the subsystems $\cal A$ and $\cal B$ is in general the smallest for the bipartition $a_2$, which is the only bipartition for the billiard B1 that contains a vertical cut.
Whether this effect is related (or not) to a suppression of certain subleading terms in the average eigenstate entanglement entropy is an open question for future work.

Summarizing the analysis of the average eigenstate entanglement entropies, we interpret the results for both tight-binding billiards as being consistent with the volume-law contributions given by Eqs.~(\ref{def_S_rmt}) and~(\ref{def_S2_rmt}).
The choice of a billiard bipartition, however, may influence the subleading terms.
The $\propto V^{-1/2}$ scaling of the volume-law coefficients in Fig.~\ref{fig6} suggest that the first subleading term scales as $v_1 V^{1/2}$, where the coefficient $v_1$ depends on a particular bipartition.

\section{Single-particle eigenstate thermalization} \label{sec:observables}

We now complement the analysis of single-particle spectral statistics from Sec.~\ref{sec:spectrum} and the entanglement content of many-body eigenstates from Sec.~\ref{sec:entanglement} with the analysis of matrix elements of observables.
We focus on the matrix elements in single-particle eigenstates of the Hamiltonian in Eq.~(\ref{eq:billiard}).
Recently, Ref.~\cite{lydzba_zhang_21} studied the matrix elements in single-particle eigenstates of the two quantum-chaotic quadratic Hamiltonians, the Dirac SYK2 Hamiltonian and the 3D Anderson Hamiltonian at weak disorder.
It was observed that the matrix elements can be described by an ansatz that shares similarities with the well-known ansatz of eigenstate thermalization hypothesis (ETH)~\cite{deutsch_91, srednicki_94, srednicki_99, rigol_dunjko_08, dalessio_kafri_16}.
The latter was predominantly studied in many-body eigenstates of interacting systems (see, e.g., Refs.~\cite{rigol_dunjko_08, rigol_09a, steinigeweg_herbrych_13, beugeling_moessner_14, sorg14, steinigeweg_khodja_14, kim_ikeda_14, mondaini_fratus_16, mondaini_rigol_17, yoshizawa_iyoda_18, jansen_stolpp_19, leblond_mallayya_19, mierzejewski_vidmar_20, brenes_leblond_20, leblond_rigol_20, richter_dymarsky_20, noh_21, sugimoto_hamazaki_21, schoenle_jansen_21}).
To highlight the single-particle nature of the studied matrix elements in quantum-chaotic quadratic Hamiltonians, Ref.~\cite{lydzba_zhang_21} dubbed this phenomenon as {\it single-particle} ETH.

The ansatz for the matrix elements of observables in the single-particle ETH can be expressed as~\cite{lydzba_zhang_21}
\begin{equation} \label{def_eth_ansatz}
\langle \alpha |\hat O |\beta\rangle = {\cal O}(\bar E) \delta_{\alpha\beta} + \rho(\bar E)^{-1/2} {\cal F}(\bar E, \omega) R_{\alpha\beta}\;,
\end{equation}
where $\bar E = (E_\alpha + E_\beta)/2$ is the mean energy, $\omega = E_\beta-E_\alpha$ is the energy difference, and ${\cal O}(\bar E)$, ${\cal F}(\bar E, \omega)$ are smooth functions of their arguments.
The single-particle density of states at energy $\bar E$ is defined as $\rho(\bar E) = \delta N/\delta E|_{\bar E}$, and $R_{\alpha\beta}$ is a random number with zero mean and unit variance.

The main goal of this section is to explore to which extent does Eq.~(\ref{def_eth_ansatz}) describe the matrix elements of observables in tight-binding billiards.
Equation~(\ref{def_eth_ansatz}) was introduced with having in mind systems without degeneracies in the single-particle spectrum and divergences in the density of states $\rho(\bar E)$.
In the tight-binding billiards studied here we only study the validity of Eq.~(\ref{def_eth_ansatz}) away from zero modes, and interpret $\rho(\bar E)$ as a quantity that increases as $\propto V$, as in other quantum-chaotic quadratic Hamiltonians studied so far.

Due to the confinement of zero modes to one of the sublattices, see Sec.~\ref{sec:zeromodes}, they are expected to exhibit some degree of nonergodicity.
A related question not addressed here is whether zero modes in tight-binding billiards may be referred to as quantum single-particle scars, i.e., states that violate the single-particle ETH.
Drawing the analogies between the wavefunction amplitudes of tight-binding billiards (see Figs.~\ref{fig14} and~\ref{fig13}) and continuum billiards (see, e.g., recent results for scars in triangular billiards in Ref.~\cite{lozej_casati_22}), one observes that they are all spatially confined to a certain fraction of the space (or lattice).
Hence one is indeed tempted to associate the zero modes of tight-binding billiards with quantum scars.
On the other hand, recent work in the context of quantum many-body scars in interacting lattice models (see, e.g., Refs.~\cite{serbyn_abanin_21, moudgalya_bernevig_22} for reviews on quantum many-body scars) have established a view that while zero modes may represent an important ingredient for the emergence of scars, only a small portion of zero modes may actually represent true scars (as defined by the absence of volume-law entanglement)~\cite{lin_motrunich_19, banerjeee_sen_21, volker_serbyn_21, biswas_banerjee_22}.
Moreover, there is in general no unique way to chose eigenfunctions in the degenerate subspace.
Since the focus of this work is to test validity of Eq.~(\ref{def_eth_ansatz}) for single-particle eigenstates at nonzero energies, we leave the analysis of statistical properties of matrix elements of zero modes to future work.

We study two local one-body observables in single-particle eigenstates.
The first is the site occupation
\begin{equation} \label{def_ni}
\hat n_i = \frac{1}{\sqrt{V-1}}(V \hat c_i^\dagger \hat c_i  - 1)\,,
\end{equation}
where we fix $i= (i_x^*,i_y^*)$ to the center of the lattice and we simplify the notation $\hat n_i \to \hat n$ further on.
The second is the next-nearest neighbor correlation 
\begin{equation} \label{def_hij}
\hat{h}_{ij} = \sqrt{\frac{V}{2}} \left( \hat{c}_{i}^{\dagger}\hat{c}^{}_{j}+\hat{c}_{j}^{\dagger}\hat{c}^{}_{i} \right)\,,
\end{equation}
where we fix $i= (i_x^*, i_y^*)$, $j = (i_x^* +1, i_y^*+1)$ and we simplify the notation $\hat h_{ij} \to \hat h$ further on.
Both observables are traceless and normalized, i.e., their Hilbert-Schmidt norm in the single-particle space is
$||\hat O||^2 \equiv \frac{1}{V}{\rm Tr}\{ \hat O^2 \}=1$~\cite{lydzba_zhang_21}.

In what follows we explore two key properties of the single-particle ETH:
fluctuations of matrix elements in Sec.~\ref{sec:fluctuations} and distributions of matrix elements in Sec.~\ref{sec:distributions}.

\subsection{Fluctuations of matrix elements} \label{sec:fluctuations}

\begin{figure}[!t]
\centering
\includegraphics[width=0.98\columnwidth]{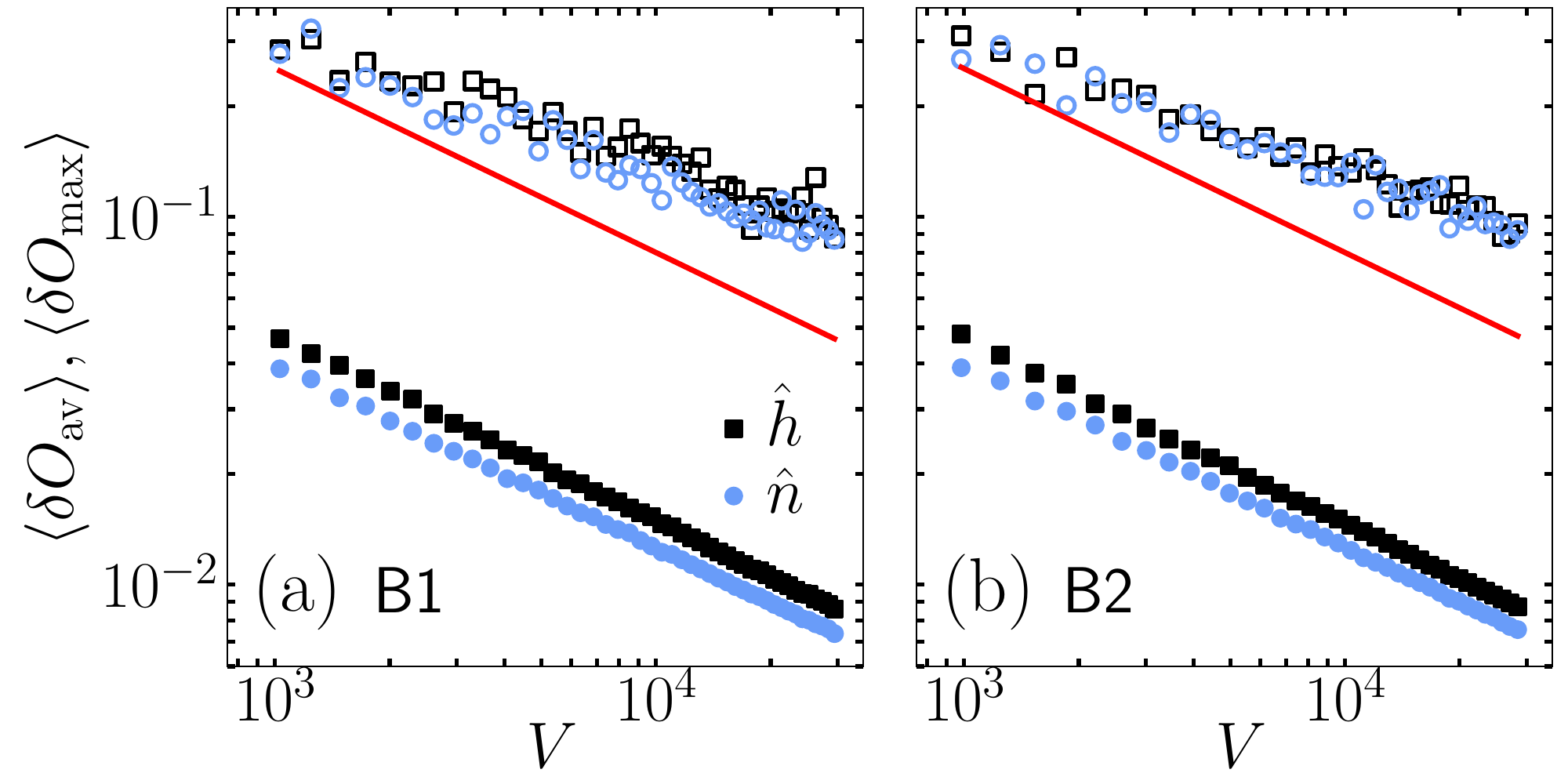}
\caption{
Eigenstate-to-eigenstate fluctuations $\langle\delta O_\text{av}\rangle$ (filled symbols) and $\langle\delta O_\text{max}\rangle$ (open symbols), see Eqs.~(\ref{dO_av}) and~(\ref{dO_max}) and the text below, as a function of $V$ in the billiards (a) B1 and (b) B2.
Results are shown from the observables $\hat{O} = \hat{n}$ (blue circles) and $\hat{O} = \hat{h}$ (black squares) from Eqs.~(\ref{def_ni}) and~(\ref{def_hij}), respectively.
The red line is proportional to $V^{-1/2}$ and is a guide to the eye.
A fitting analysis of the results is carried out in Fig.~\ref{fig9}.
}\label{fig8}
\end{figure}

We first study the eigenstate-to-eigenstate fluctuations of the diagonal matrix elements of observables.
Using the notation $O_{\alpha\beta} \equiv \langle\alpha|\hat O|\beta\rangle$, the observable fluctuations between two consecutive eigenstates are
$\delta O_\alpha = O_{\alpha\alpha} - O_{\alpha-1,\alpha-1}$.
We then study the mean eigenstate-to-eigenstate fluctuations,
\begin{equation} \label{dO_av}
\delta O_\text{av}=||\Lambda||^{-1} \sum_{\ket{\alpha}\in\Lambda} |\delta O_\alpha| \,,
\end{equation}
where $\Lambda$ is a set of states $|\alpha\rangle$ that comprise $80\%$ of eigenstates in the middle of the spectrum excluding zero modes, i.e., $||\Lambda||$ is slightly smaller than $0.8\, V$.
We also calculate the maximal eigenstate-to-eigenstate fluctuations over the same set of eigenstates defined as
\begin{equation} \label{dO_max}
\delta O_\text{max} = \text{max}_{\ket{\alpha}\in\Lambda}|\delta O_\alpha| \,.
\end{equation}
In the actual numerical calculations we average $\delta O_\text{av}$ and $\delta O_\text{max}$ over 5 billiards with similar number of lattice sites, and plot them versus the mean number of lattice sites of these billiards.
We denote the corresponding averages as $\langle\delta O_\text{av}\rangle$ and $\langle\delta O_\text{max}\rangle$.

The measures from Eqs.~(\ref{dO_av}) and~(\ref{dO_max}) were first introduced in studies of ETH in interacting systems~\cite{kim_ikeda_14}.
In particular, vanishing of the maximal differences $\delta O_\text{max}$~(\ref{dO_max}) with increasing the system size has now become one of the defining measures of the validity of the ETH.
It was shown for various interacting (non-integrable) models that in the bulk of the spectrum, $\delta O_\text{max}$ vanishes exponentially fast with the number of lattice sites~\cite{mondaini_fratus_16, luitz_16, jansen_stolpp_19}.
The exponential dependence stems from the scaling of the density of states in the ETH ansatz, i.e., by replacing $\rho(\bar E)$ in Eq.~(\ref{def_eth_ansatz}) with the many-body density of states.
Recently, it was argued that the corresponding scaling of $\delta O_\text{av}$ and $\delta O_\text{max}$  for the single-particle ETH is polynomial in $V$, namely $\propto V^{-\zeta}$, with $\zeta \lesssim 0.5$~\cite{lydzba_zhang_21}.

\begin{figure}[!t]
\centering
\includegraphics[width=0.98\columnwidth]{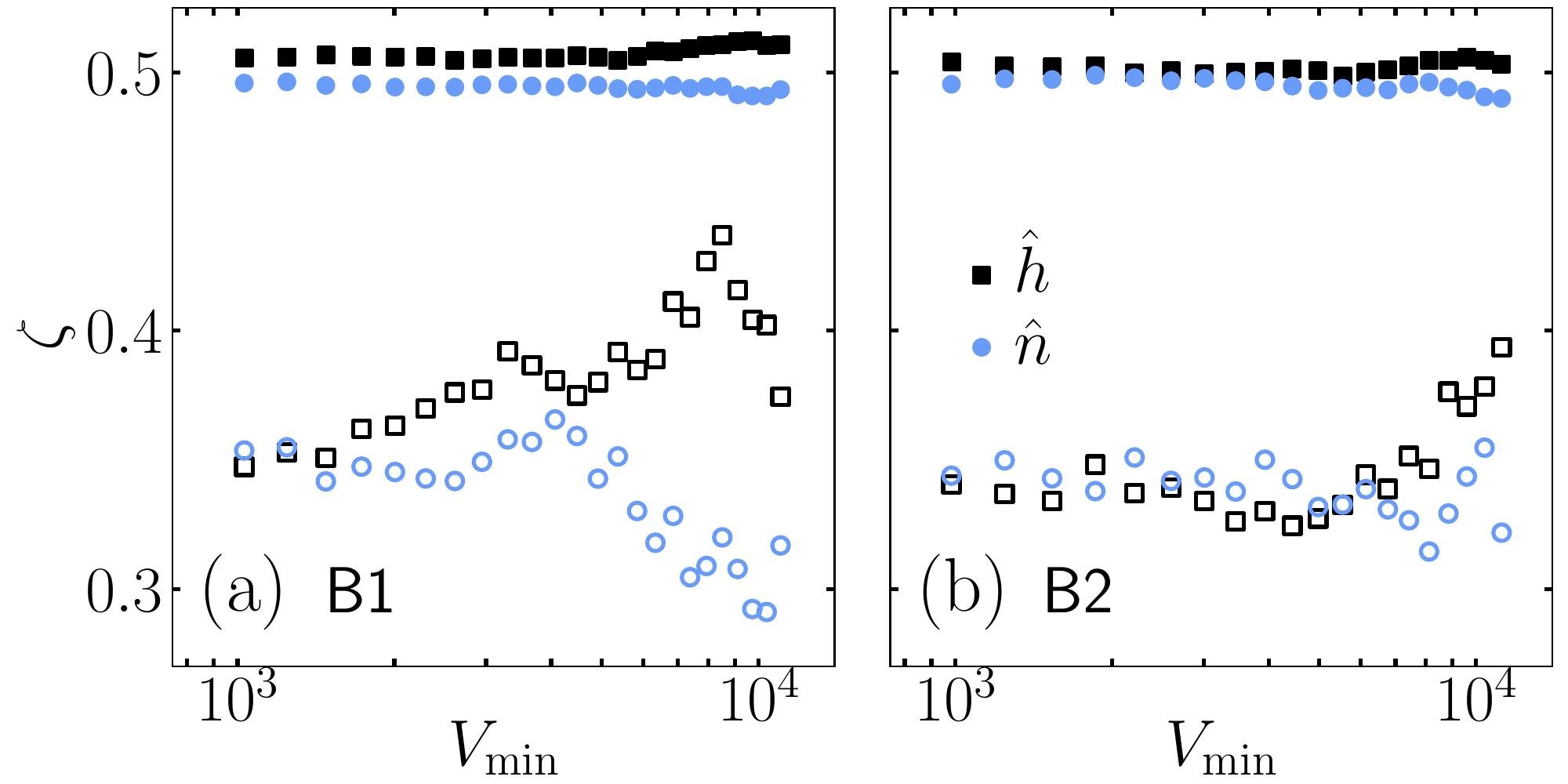}
\caption{
Exponents $\zeta$ of the power-law fits $aV^{-\zeta}$ to the results in Fig.~\ref{fig8}.
We show $\zeta$ as a function of $V_{\rm min}$, where the latter denotes the smallest $V$ of the results included in the fit (the largest $V$ always corresponds to the largest system that we studied).
Filled symbols denote fits to the results for $\langle\delta O_\text{av}\rangle$ vs $V$ and open symbols to those for $\langle\delta O_\text{max}\rangle$ vs $V$.
Results are shown for the observables $\hat{O} = \hat{n}$ (blue circles) and $\hat{O} = \hat{h}$ (black squares) from Eqs.~(\ref{def_ni}) and~(\ref{def_hij}), respectively, in the billiards (a) B1 and (b) B2.
}\label{fig9}
\end{figure}

Figure~\ref{fig8} shows the scalings of $\langle\delta O_\text{av}\rangle$ and $\langle\delta O_\text{max}\rangle$ with $V$ for the observables from Eqs.~(\ref{def_ni}) and~(\ref{def_hij}).
Both measures appear to decrease to zero in the thermodynamic limit $V \to \infty$.
The average fluctuations $\langle\delta O_\text{av}\rangle$ are rather small for the system sizes under investigation and their decrease is consistent with the functional form $\propto V^{-1/2}$, as expected from the single-particle ETH.
While the values of maximal fluctuations $\langle\delta O_\text{max}\rangle$ are roughly an order of magnitude larger than those for $\langle\delta O_\text{av}\rangle$, they also appear to exhibit a slower decrease with $V$ than those for $\langle\delta O_\text{av}\rangle$.

A quantitative analysis of the decay of fluctuations with $V$ is carried out in Fig.~\ref{fig9}.
Specifically, we first perform a power-law fit $a V^{-\zeta}$ to the results in Fig.~\ref{fig8}, and then show the exponent $\zeta$ as a function of $V_{\rm min}$ in Fig.~\ref{fig9}, where $V_{\rm min}$ denotes the smallest system size included in the fit.
Results confirm that $\zeta \approx 0.50$ for the average fluctuations $\langle\delta O_\text{av}\rangle$.
On the other hand, we observe approximately $\zeta \in [0.3,0.4]$ for the maximal fluctuations $\langle\delta O_\text{max}\rangle$, which is consistent with the results in the 3D Anderson model at weak disorder~\cite{lydzba_zhang_21}.
The question of whether $\zeta$ approaches $1/2$ in the limit $V_{\rm min} \to \infty$ (as it does for the Dirac SYK2 model~\cite{lydzba_zhang_21}) is an interesting question, however, it is beyond the scope of this work.

\begin{figure}[!t]
\centering
\includegraphics[width=0.98\columnwidth]{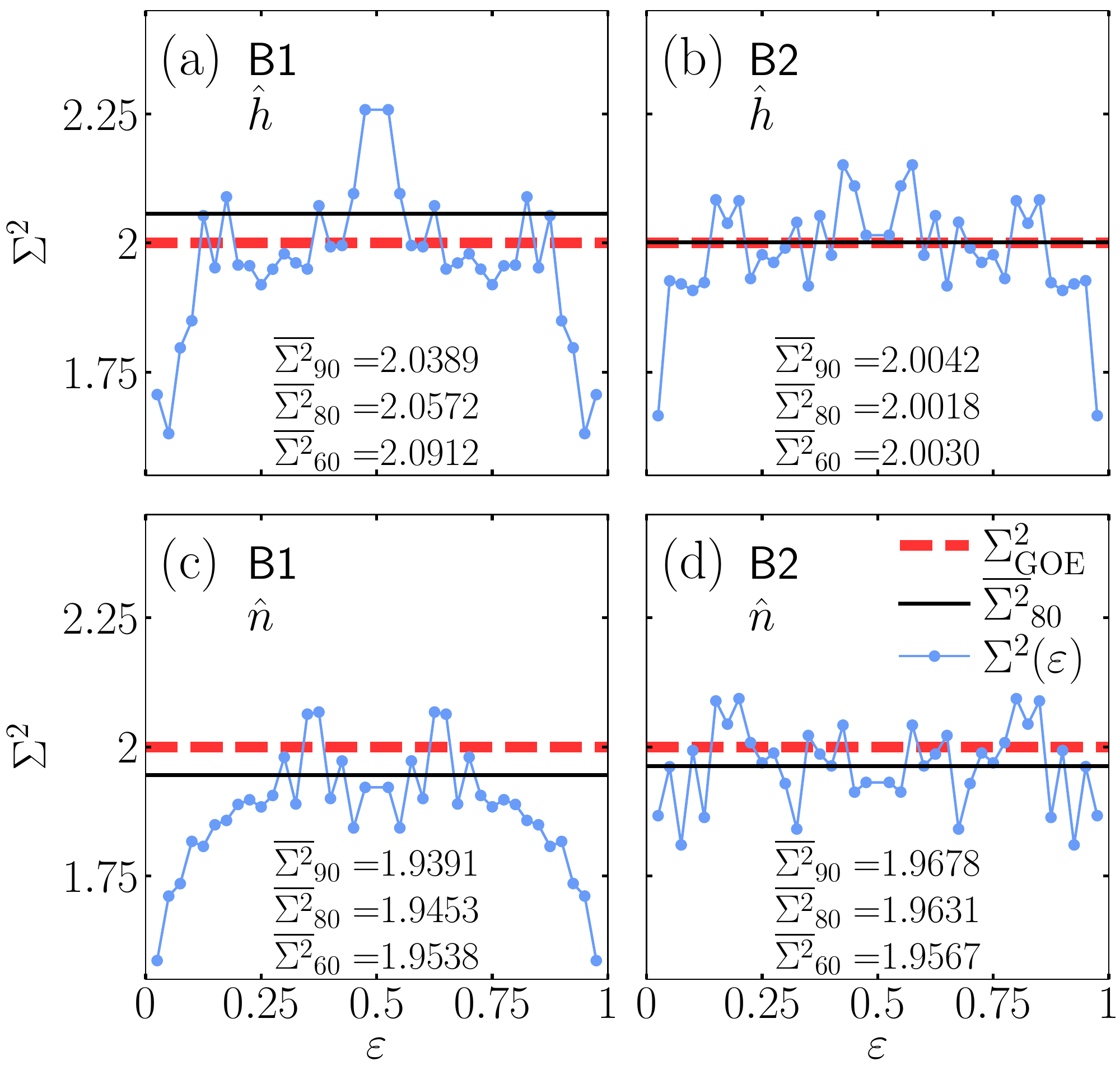}
\caption{
Ratio of variances $\Sigma^2$ from Eq.~(\ref{def_sigma2}) versus normalized energy $\varepsilon = (E-E_{\rm min})/(E_{\mathrm{max}} - E_{\mathrm{min}})$.
Results are shown for the observables (a, b) $\hat{h}$ and (c, d) $\hat{n}$,
in the billiards (a, c) B1 and (b, d) B2.
Blue dots are results for $\Sigma^2(\varepsilon)$, where the variances in Eqs.~(\ref{def_variance_diag}) and~(\ref{def_variance_off}) are calculated at the target eigenenergies that are closest to the corresponding $\varepsilon$.
The values of $\Sigma^2(\varepsilon)$ are further averaged over $20$ system sizes spanning from approximately $V = 25000$ to $V = 30000$.
We plot $\Sigma^2$ for $39$ equally spaced values of $\varepsilon$ and leave out the middle point at $\varepsilon = 0.5$, which is the only point containing zero modes.
The red dashed line represents the GOE value $\Sigma_{\mathrm{GOE}}^2 = 2$.
In the legends we list the average values $\overline{\Sigma^2}_{90}$, $\overline{\Sigma^2}_{80}$ and $\overline{\Sigma^2}_{60}$.
They are obtained by first calculating $\Sigma^2$ at all eigenenergies (except for 100 states at the spectral edges), and then averaging the results over $90\%, 80\%$ and $60\%$ eigenenergies, respectively, in the middle of the spectrum (excluding zero modes), and over 20 system sizes.
The black solid line represents $\overline{\Sigma^2}_{80}$.
}\label{fig7}
\end{figure}

Another measure of fluctuations is related to the variances of matrix elements in certain energy windows.
The variance of the diagonal matrix elements is
\begin{equation} \label{def_variance_diag}
    \sigma^2_\text{diag}=||\Lambda||^{-1} \sum_{\ket{\alpha}\in\Gamma} O^2_{\alpha\alpha}-
    \left(||\Lambda||^{-1}\sum_{\ket{\alpha}\in\Gamma} O_{\alpha\alpha}\right)^2 \,,
\end{equation}
where $\Lambda$ is a set of 201 eigenstates ($||\Lambda|| = 201$) around the target energy $E$. Analogously, the variance of the off-diagonal matrix elements is
\begin{equation} \label{def_variance_off}
    \sigma^2_\text{off}=||\Lambda'||^{-1} \sum_{\substack{\ket{\alpha},\ket{\beta}\in\Lambda \\ \ket{\alpha}\neq\ket{\beta}}} O^2_{\alpha\beta}
    -\left(||\Lambda'||^{-1} 
    \sum_{\substack{\ket{\alpha},\ket{\beta}\in\Lambda \\ \ket{\alpha}\neq\ket{\beta}}} O_{\alpha\beta}\right)^2 \,,
\end{equation}
where $||\Lambda'|| = ||\Lambda||^2-||\Lambda||=40200$.
A hallmark of the ETH is that $\sigma^2_\text{diag}$ decays as the inverse of the Hilbert-space dimension.
In case of the single-particle ETH, one expects $\sigma^2_\text{diag} = a_1/V$~\cite{lydzba_zhang_21}, where $a_1$ is a constant and $V$ is the dimension of the single-particle Hilbert space.
This is consistent with the scaling $\langle\delta O_\text{av}\rangle \propto V^{-1/2}$ found in Figs.~\ref{fig8} and~\ref{fig9}.
On the other hand the decay of the variance $\sigma^2_\text{off} = a_2/V$ (or, more generally, the decay with the inverse Hilbert space dimension), where $a_2$ is a constant, may not be unique to systems complying with ETH, but could also be found in other systems such as integrable interacting models~\cite{leblond_mallayya_19, zhang_vidmar_22}.
In the Dirac SYK2 model one has $a_1=2$ and $a_2 = 1$~\cite{lydzba_zhang_21, dalessio_kafri_16}.
In other quantum-chaotic systems $a_1$ and $a_2$ may take arbitrary values, however, their ratio is expected to obey predictions from the GOE of the random matrix theory, $a_1/a_2 = 2$~\cite{mondaini_rigol_17, jansen_stolpp_19, schoenle_jansen_21}.
We hence test this prediction by defining the ratio of variances as
\begin{equation} \label{def_sigma2}
    \Sigma^2 = \frac{\sigma_{\rm diag}^2}{\sigma_{\rm off}^2} \;,
\end{equation}
for which the GOE result is $\Sigma_{\rm GOE}^2 = 2$~\cite{dalessio_kafri_16}.

In Fig.~\ref{fig7} we show the ratio $\Sigma^2$ of the observables $\hat{h}$ and $\hat{n}$ in the billiards B1 and B2.
We plot the results as a function of normalized energy $\varepsilon = (E-E_{\rm min})/(E_{\mathrm{max}} - E_{\mathrm{min}})$, thereby exploring the entire energy range from the vicinity of a ground state to highly excited states (excluding zero modes).
Due to the reflection symmetry of the energy spectrum, and hence similarity of the coefficients of the eigenstates $|\alpha\rangle$ and $|\beta\rangle$ for which $E_\beta = - E_\alpha$, see Eqs.~(\ref{def_alpha}) and~(\ref{def_beta}), the variances of the observables $\hat{h}$ and $\hat{n}$ are symmetric w.r.t.~the middle of the spectrum at $\varepsilon=0.5$.
While this is an exact statement for $\sigma_{\rm diag}^2$ in Eq.~(\ref{def_variance_diag}) and the mean of $O_{\alpha\beta}^2$ in Eq.~(\ref{def_variance_off}), we note that the values of $O_{\alpha\beta}$ [contributing to the second term on the r.h.s.~of Eq.~(\ref{def_variance_off})] may be subject to a random global change of sign of the wavefunctions.
However, since the contribution of the mean of $O_{\alpha\beta}$ to $\sigma_{\rm off}^2$ in Eq.~(\ref{def_variance_off}) is vanishingly small, the ratio of variances $\Sigma^2$ in Fig.~\ref{fig7} appears to be perfectly symmetric around $\varepsilon=0.5$.

Numerical results for the averages in Fig.~\ref{fig7} are consistent with the GOE prediction $\Sigma_{\rm GOE}^2 = 2$.
The absolute differences are small, they are $O(10^{-2})$, while in the case for observable $\hat h$ and the billiard B2 the difference is $O(10^{-3})$.
This level of agreement is comparable to the most accurate studies of the ETH in quantum-chaotic interacting systems~\cite{mondaini_rigol_17, jansen_stolpp_19, schoenle_jansen_21}.
Results in Fig.~\ref{fig7} also show some fluctuations above the average, and in some cases also a trend to lower values of $\Sigma^2$ at the spectral edges.

\subsection{Distributions of matrix elements} \label{sec:distributions}

\begin{figure}[!t]
\centering
\includegraphics[width=0.98\columnwidth]{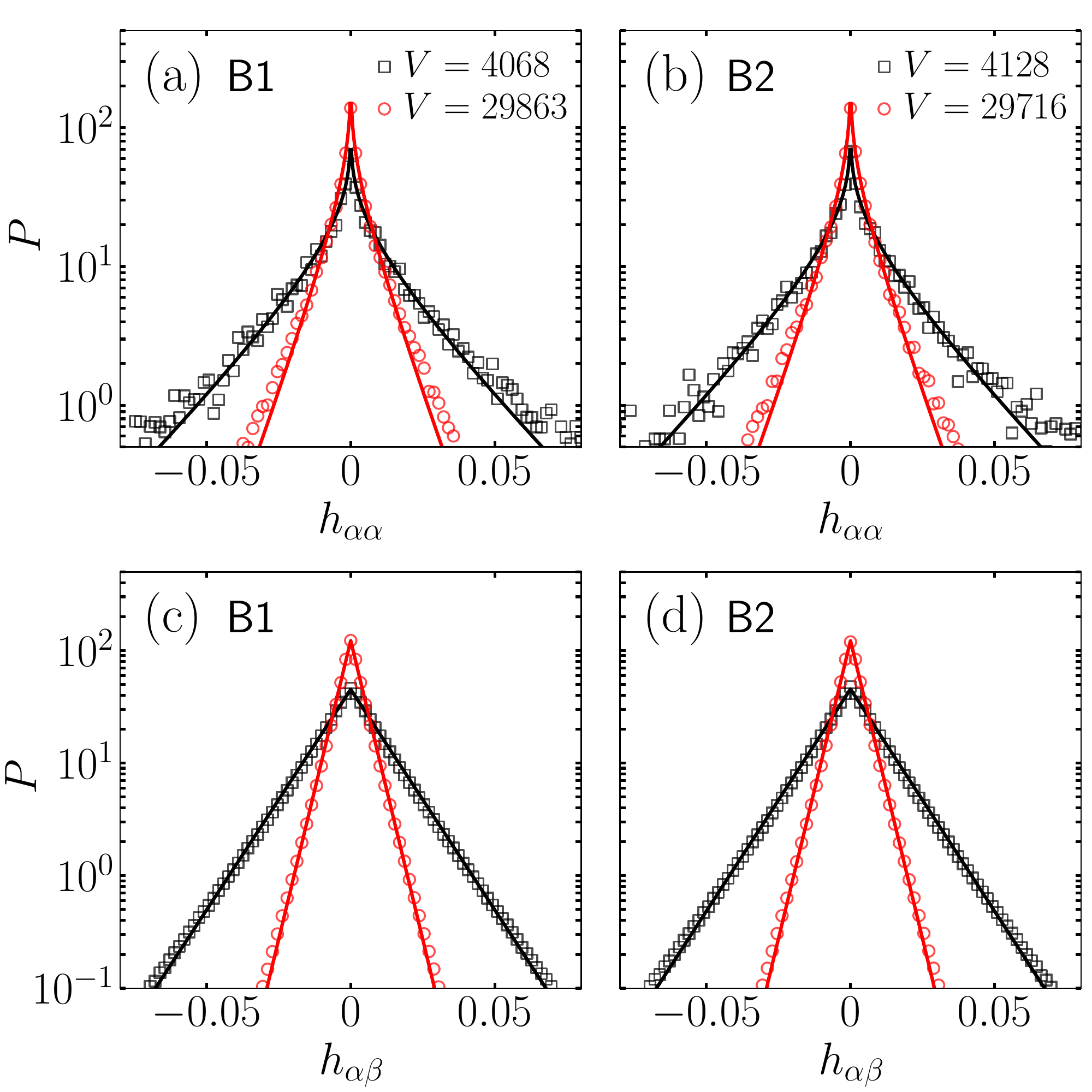}
\caption{
Probability density function $P$ of the matrix elements of $\hat h$.
Results are shown for (a, b) diagonal matrix elements and (c, d) off-diagonal matrix elements, in the billiards (a, c) B1 and (b, d) B2.
We consider matrix elements from the entire spectrum (excluding zero modes), and further average the results over $5$ systems with similar volume that is close to the mean volume $V$.
Black squares are results for smaller systems with $V \approx 4000$, red circles are results for larger systems with $V \approx 30000$.
Solid lines are the PDFs from (a, b) Eq.~(\ref{def_pdf_haa}) and (c, d) Eq.~(\ref{def_pdf_hab}).
}\label{fig10}
\end{figure}

\begin{figure}[!t]
\centering
\includegraphics[width=0.98\columnwidth]{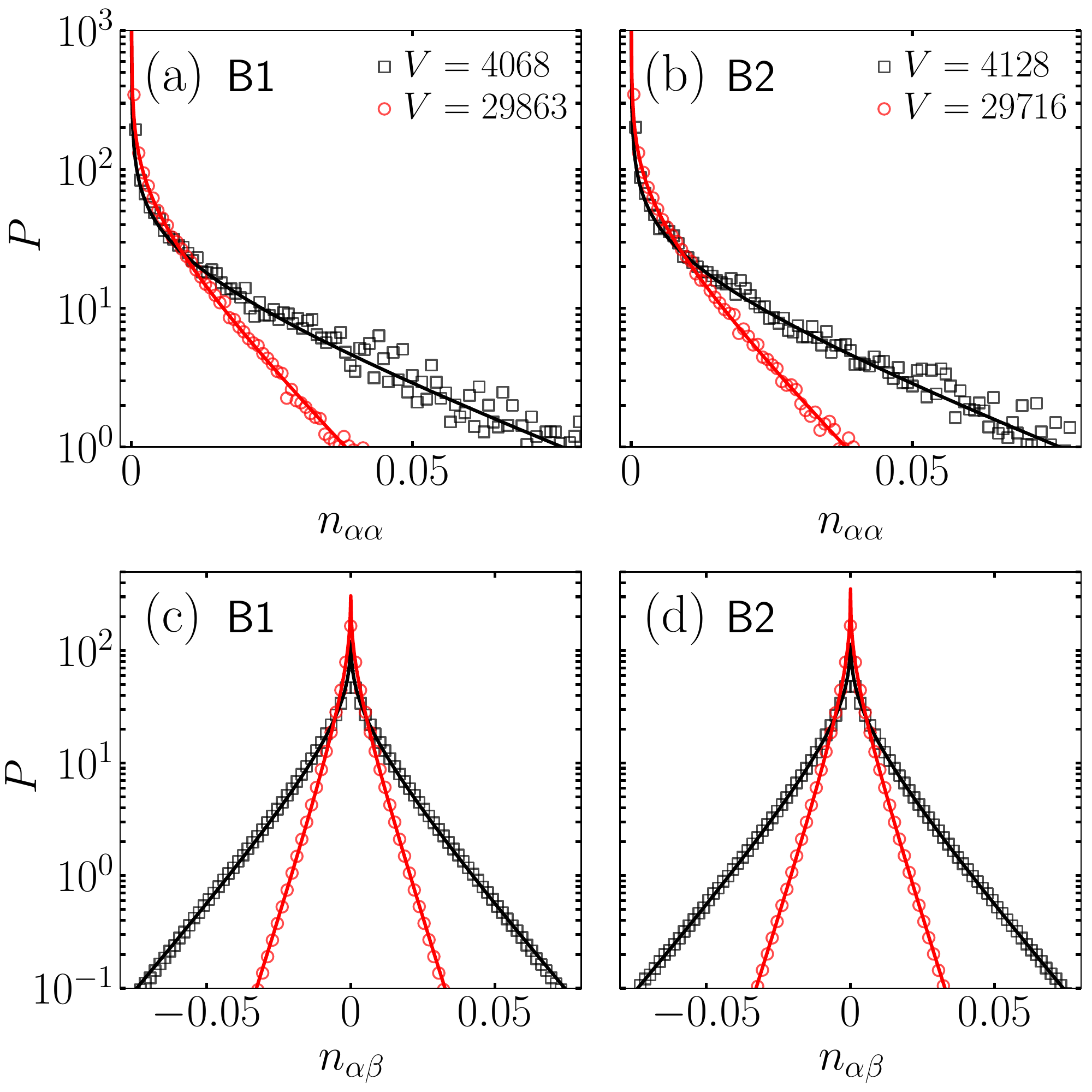}
\caption{
Probability density function $P$ of the matrix elements of $\hat n$.
Results are shown for (a, b) diagonal matrix elements and (c, d) off-diagonal matrix elements, in the billiards (a, c) B1 and (b, d) B2.
We consider matrix elements from the entire spectrum (excluding zero modes), and further average the results over $5$ systems with similar volume that is close to the mean volume $V$.
Black squares are results for smaller systems with $V \approx 4000$, red circles are results for larger systems with $V \approx 30000$.
Solid lines are the PDFs from (a, b) Eq.~(\ref{def_pdf_naa}) and (c, d) Eq.~(\ref{def_pdf_nab}).
The results in (a, b) are shifted in the $x$-axis by $1/\sqrt{V}$, such that the shifted matrix elements are non-negative.
}\label{fig11}
\end{figure}

Finally, we study the distributions of matrix elements of the observables $\hat h$ and $\hat n$.
We ask whether these distributions in the tight-binding billiards share properties with the distributions in other quantum-chaotic quadratic models.
Ref.~\cite{lydzba_zhang_21} observed that the distribution of local observables in single-particle eigenstates of quantum-chaotic quadratic models such as the 3D Anderson model and the Dirac SYK2 model may not be Gaussian.
This property for local observables appears to be unique to quantum-chaotic quadratic models, since in quantum-chaotic interacting models the distributions were always found to be Gaussian~\cite{beugeling_moessner_15, luitz_barlev_16, khaymovich_haque_19, leblond_mallayya_19, brenes_leblond_20, brenes_goold_20, leblond_rigol_20, santos_perezbernal_20, noh_21, brenes_pappalardi_21}.
We note that non-Gaussian distributions in quantum-chaotic interacting models were reported for non-local operators with diverging Hilbert-Schmidt norm, dubbed behemoths in~\cite{khaymovich_haque_19}.

Numerical results for the probability density functions (PDFs) of the matrix elements of $\hat h$ (and $\hat n$) are shown as symbols in Fig.~\ref{fig10} (and Fig.~\ref{fig11}).
The PDF, $P$, of a variable $x$ in an interval $[x,x+\Delta x]$ is defined as
\begin{equation}
    P(x) = \frac{1}{\cal N}\frac{\Delta {\cal N}}{\Delta x} \;,
\end{equation}
where $\cal N$ is the total number of elements.
The numerical PDFs shown in Figs.~\ref{fig10} and~\ref{fig11} are calculated for matrix elements in the entire spectrum (excluding zero modes), and are further averaged over 5 systems with volumes close to the mean volume $V$.
The PDFs are shown in Figs.~\ref{fig10} and~\ref{fig11} for two mean volumes $V$.
The width of the PDFs shrinks with $V$, which is a consequence of the vanishing matrix elements fluctuations, as discussed in Sec.~\ref{sec:fluctuations}.
Most importantly, all the PDFs appear to be non-Gaussian.

We contrast the numerical results to the analytical predictions.
The latter are obtained assuming that the coefficients $\{u_i^{(\alpha)}\}$, $\{ v_j^{(\alpha)} \}$ in single-particle eigenstates $\{|\alpha\rangle\}$, see Eq.~(\ref{def_alpha}), behave as random variables drawn from a normal distribution with zero mean and variance $1/V$.
This assumption carries similarities with the Berry conjecture about the structure of chaotic wavefunctions~\cite{berry_77}.
Below we list the resulting analytical expressions of the distributions for the observables under investigation, while a complete derivation of these expressions can be found in~\cite{lydzba_zhang_21}.

The PDF of diagonal matrix elements of $\hat h$ from Eq.~(\ref{def_hij}) is
\begin{equation} \label{def_pdf_haa}
P_{h_{\alpha\alpha}} (x) = \frac{1}{\pi}\sqrt{\frac{V}{2}}K_0\left(\sqrt{\frac{V}{2}}|x|\right) \,,
\end{equation}
where $K_0$ is a modified Bessel function of the second kind.
The latter emerges since $P_{h_{\alpha\alpha}}$ is approximated as a product distribution of normal random variables.
The PDF of off-diagonal matrix elements of $\hat h$ is obtained from a sum distribution, and yields the exponential distribution,
\begin{equation} \label{def_pdf_hab}
P_{h_{\alpha\beta}} (x) = \sqrt{\frac{V}{2}} e^{-\sqrt{2V}|x|} \,.
\end{equation}
The PDF of diagonal matrix elements of $\hat n$ from Eq.~(\ref{def_ni}) is related to that of the square of normal random variables, and it is described by a chi-square distribution with degree 1,
\begin{equation} \label{def_pdf_naa}
P_{n_{\alpha\alpha}} (x) =\frac{V^{1/4}}{\sqrt{2\pi}} \frac{1}{\sqrt{x+\frac{1}{\sqrt{V}}}} e^{-\frac{\sqrt{V}}{2}\left[x+\frac{1}{\sqrt{V}}\right]} \,.
\end{equation}
The PDF of off-diagonal matrix elements of $\hat n$ is related to that of the product distribution of normal random variables and is, up to normalization, identical to the PDF of diagonal matrix elements of $\hat h$.
It is given by
\begin{equation} \label{def_pdf_nab}
P_{n_{\alpha\beta}} (x) = \frac{\sqrt{V}}{\pi}\text{K}_{0}\left(\sqrt{V}|x|\right) \,,
\end{equation}
where $K_0$ is again a modified Bessel function of the second kind.

The analytical expressions from Eqs.~(\ref{def_pdf_haa})-(\ref{def_pdf_nab}) are shown as lines in Figs.~\ref{fig10}-\ref{fig11}.
They very accurately describe the numerical results, which are shown as symbols.
The agreement suggest that the overwhelming majority of the matrix elements of local observables in the tight-binding billiards under investigation comply with those in other quantum-chaotic quadratic models, and are well described by the single-particle ETH.

\section{Conclusions} \label{sec:conclusions}

In this work we explored the fate of quantum-chaotic quadratic Hamiltonians in the absence of disorder.
To this end we introduced the tight-binding billiards, i.e., the systems of free fermions on square lattices with curved boundaries.
Even though these systems share some similarities with billiards in continuum, the lattice discretization introduces an ultra-high energy scale to the tight-binding billiards such that a direct quantitative comparison does not seem to be obvious.
In fact, it appears that the tight-binding billiards may represent a class of quantum systems that lie in between single-particle quantum billiards in continuum and interacting many-body quantum systems in the lattice.

We showed that the tight-binding billiards exhibit several universal properties.
Most importantly, (a) the average eigenstate entanglement entropy of many-body eigenstates agrees reasonably well with predictions from random-matrix theory~\cite{lydzba_rigol_20} and with results for the typical pure fermionic Gaussian states~\cite{bianchi_hackl_21, bianchi_hackl_22}, and (b) the statistical properties of observables in non-degenerate single-particle eigenstates comply with the single-particle ETH~\cite{lydzba_zhang_21}.
These properties establish close connections of tight-binding billiards with other quantum-chaotic quadratic Hamiltonians studied in the past, i.e., the 3D Anderson model below the localization transition and the Dirac SYK2 model~\cite{lydzba_rigol_21, lydzba_zhang_21}.
As a side result, we derived a closed-form expression of the average 2nd R\' enyi entanglement entropy as a function of the subsystem fraction, which builds on previous results from Ref.~\cite{liu_chen_18, zhang_liu_20} and complements a similar expression found before for the von Neumann entanglement entropy~\cite{lydzba_rigol_20}.
Moreover, we observed a subextensive number of zero modes, i.e., degenerate single-particle eigenstates with zero energy.
We interpreted them as chiral particles and argued that their wavefunction is confined to one of the sublattices.

Our results may stimulate future studies of several intriguing properties of tight-binding billiards.
First, one should explore the validity of universal eigenstate entanglement entropy and the single-particle ETH in other billiard geometries, in particular those that are associated with a weaker degree of ergodicity in the corresponding continuum billiards.
Then it would also be interesting to explore how generic is the mechanism that gives rise to zero modes and the emergence of chiral particles.
The zero modes studied here represent a platform to sharpen the notion of quantum scars in tight-binding billiards in the future, which may give rise to some form of weak ergodicity breaking of the single-particle ETH.
Particularly interesting is the question of quantum quench dynamics and its characterization for different classes of initial states, especially those that have large overlaps with zero modes.

\acknowledgements
We acknowledge discussions with L. Hackl, M. Lek\v{s}e, \v{C}. Lozej, P. \L yd\.{z}ba, M. Mierzejewski, J. Mravlje, T. Prosen, T. Rejec, M. Rigol, and L. Ul\v cakar.
This work was supported by the the Slovenian Research Agency (ARRS), Research core fundings Grants No.~P1-0044 (I.U.~and L.V.) and No.~J1-1696 (L.V.).

\appendix

\section{Wavefunction structure of zero modes} \label{app:zeromodes}

In Sec.~\ref{sec:zeromodes} we presented some general properties of zero modes.
Here we extend these arguments to obtain information about their wavefunctions.
The analysis is built on the structure of the Hamiltonian matrix sketched in Fig.~\ref{fig_sublattice}(b) that consists of four blocks, i.e., two diagonal and two off-diagonal blocks.
The diagonal blocks of dimensions $N_A \times N_A$ and $N_B \times N_B$ are zero, while the off-diagonal blocks of dimensions $N_A \times N_B$ and $N_B \times N_A$ may include nonzero elements.

\begin{figure*}[!t]
\centering
\includegraphics[width=2.0\columnwidth]{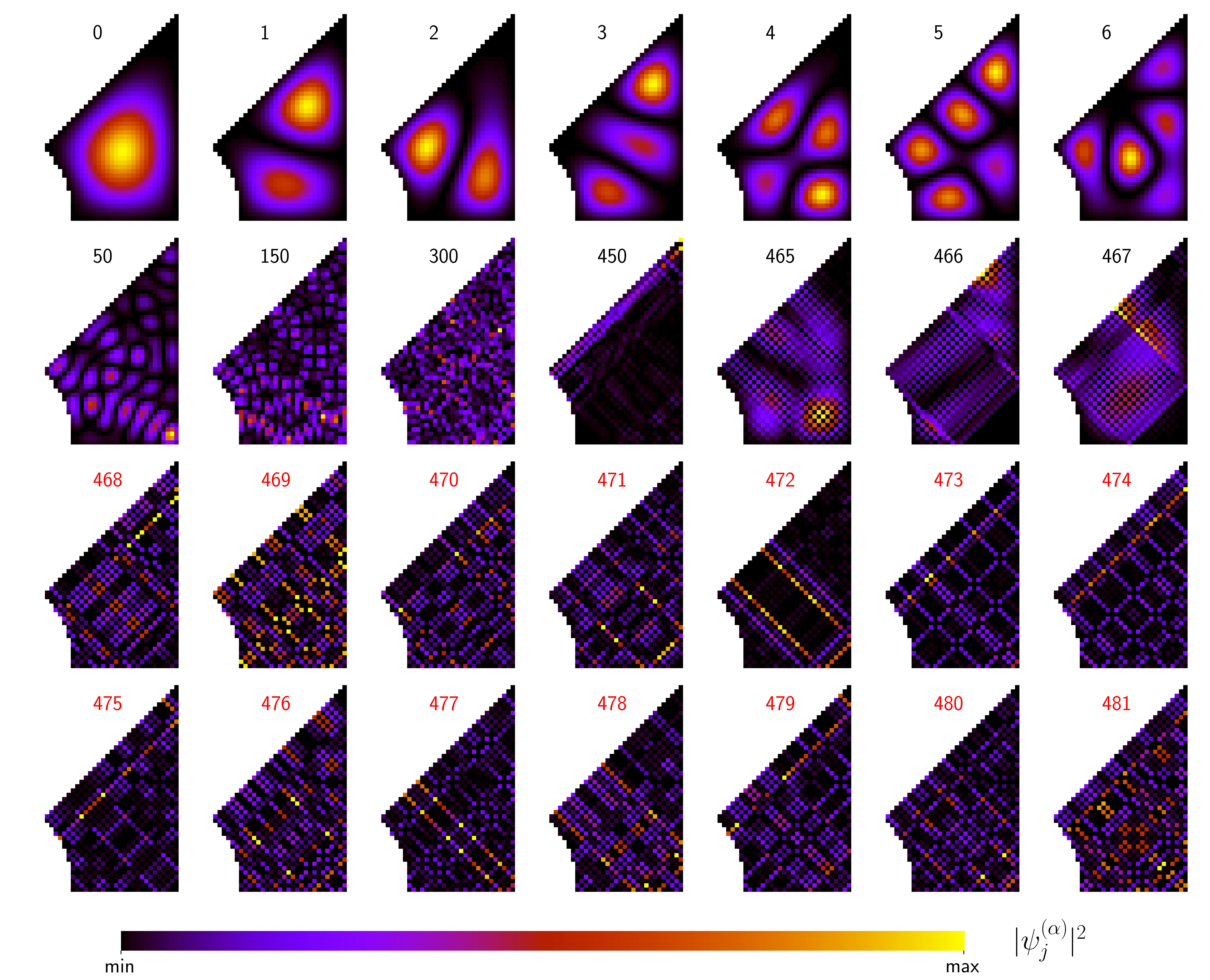}
\caption{
Amplitudes $|\psi_j^{(\alpha)}|^2$ of the Hamiltonian eigenfunctions in the single-particle site-occupation basis, $|\alpha\rangle = \sum_{j=1}^V \psi_j^{(\alpha)} |j\rangle$, where $|j\rangle \equiv \hat c_j^\dagger |\emptyset\rangle$, in the billiard B1 at $V = 950$.
Numbers correspond to the eigenfunction index $\alpha$.
Black numbers (upper two rows) refer to eigenfunctions at nonzero energy, while red numbers (lower two rows) refer to zero modes.
The color scale in the density plot is adjusted in each panel to the corresponding minimal and maximal value.
}\label{fig14}
\end{figure*}

\begin{figure*}[!t]
\centering
\includegraphics[width=2.0\columnwidth]{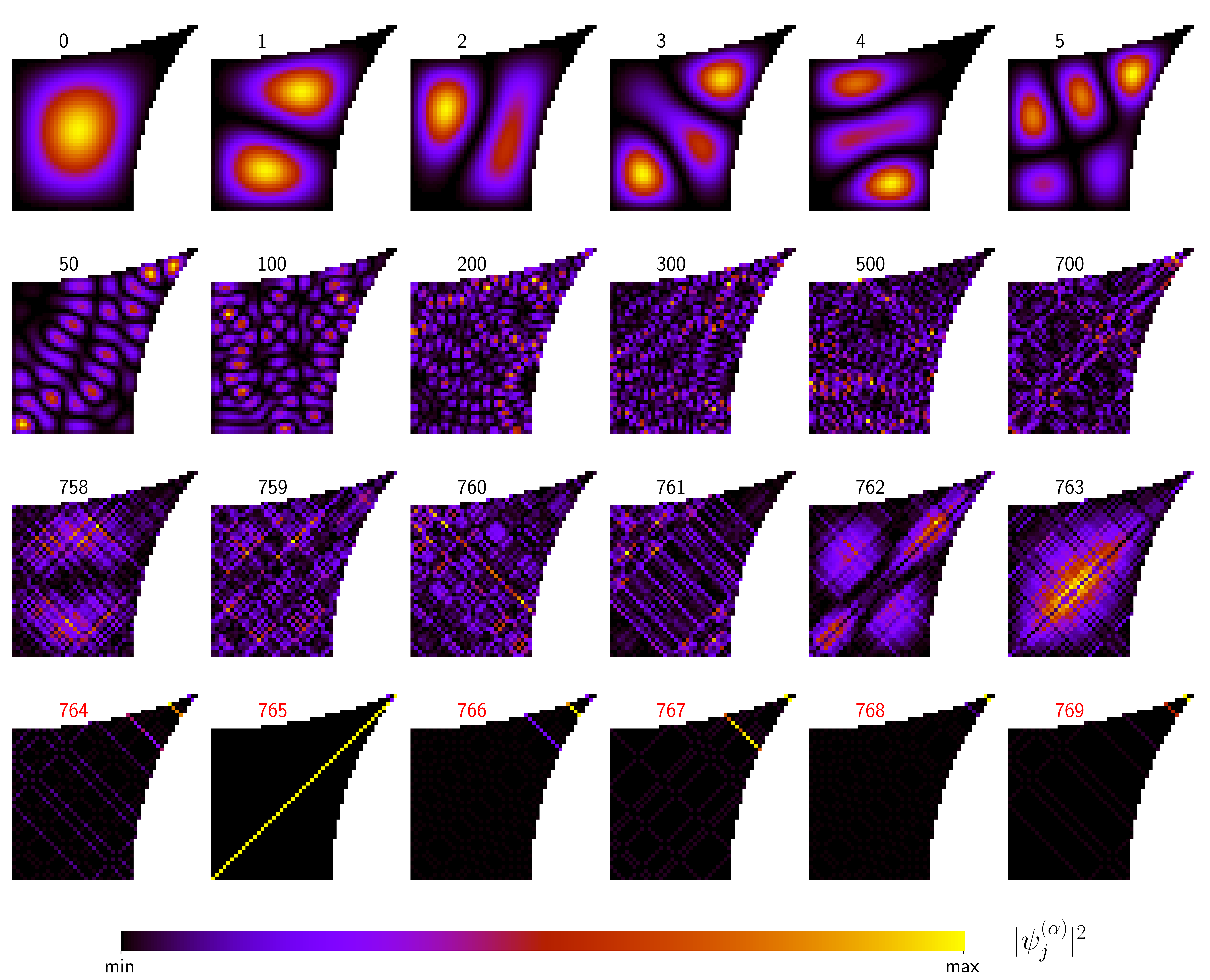}
\caption{
Amplitudes $|\psi_j^{(\alpha)}|^2$ of the Hamiltonian eigenfunctions in the single-particle site-occupation basis, $|\alpha\rangle = \sum_{j=1}^V \psi_j^{(\alpha)} |j\rangle$, where $|j\rangle \equiv \hat c_j^\dagger |\emptyset\rangle$, in the billiard B2 at $V = 1534$.
Numbers correspond to the eigenfunction index $\alpha$.
Black numbers (upper three rows) refer to eigenfunctions at nonzero energy, while red numbers (lower row) refer to zero modes.
The color scale in the density plot is adjusted in each panel to the corresponding minimal and maximal value.
}\label{fig13}
\end{figure*}

Let us first consider the case in which all the columns in the off-diagonal block of dimensions $N_A \times N_B$ [bottom left block in Fig.~\ref{fig_sublattice}(b)] are linearly independent.
This implies $m=0$ in Eq.~(\ref{def_rank_H_m}) and hence the rank of the Hamiltonian is $\mathrm{rank}(H) = 2N_B = V- \delta N$
(we assume $N_A \geq N_B$, as in the main text).
We show below that in this case, the wavefunctions of all zero modes are confined into the larger sublattice, i.e., sublattice A.

Suppose $|\alpha\rangle$ is a zero mode of the Hamiltonian. We can write $|\alpha\rangle = |\alpha_A\rangle + |\alpha_B\rangle$, where $|\alpha_A\rangle$ only includes occupations in sublattice A and $|\alpha_B\rangle$ only includes occupations in sublattice $B$. Then
\begin{equation}
\label{eq_zero}
    \hat{H}|\alpha\rangle = 0 \implies \hat{H}\left(|\alpha_A\rangle + |\alpha_B\rangle\right) = \hat{H}|\alpha_A\rangle + \hat{H}|\alpha_B\rangle = 0\,.
\end{equation}
Since $\hat{H}|\alpha_A\rangle \in \mathcal{H}_B$ and $\hat{H}|\alpha_B\rangle \in \mathcal{H}_A$, each term must equal to zero, i.e., $\hat{H}|\alpha_A\rangle = 0$ and $\hat{H}|\alpha_B\rangle = 0$
($\mathcal{H}_A$ and $\mathcal{H}_B$ are single-particle Hilbert spaces in sublattices A and B, respectively).
If we write $|\alpha_B\rangle$ as a sum over the position basis vectors in sublattice B, $|\alpha_B\rangle = \sum_{i = 1}^{N_B}v_i |b_i\rangle$, we have
\begin{align}
\hat{H}|\alpha_B\rangle & = \hat{H} \sum_{i = 1}^{N_B}v_i |b_i\rangle = \sum_{i = 1}^{N_B} v_i \hat{H}|b_i\rangle = 0\,, \label{eq_columns_1} \\
&\implies \sum_{i = 1}^{N_B} v_i |s_i\rangle = 0\,, \label{eq_columns}
\end{align}
where $|s_i\rangle$ can be represented as columns in the $N_A \times N_B$ off-diagonal block in the Hamiltonian matrix. Since all columns are linearly independent, Eq.~(\ref{eq_columns}) then requires $v_i = 0$ for $i = 1, \dots, N_B$, and as a consequence,
\begin{equation} \label{def_alphaB_zero}
    |\alpha_B\rangle = \sum_{i = 1}^{N_B}v_i |b_i\rangle = 0\;.
\end{equation}
Zero modes hence only have occupations in the larger sublattice A.

As argued in Sec.~\ref{sec:zeromodes}, the case considered above is found in the billiard B1.
In Fig.~\ref{fig14} we show examples of the wavefunction amplitudes in this billiard at $V=950$.
The upper two rows in Fig.~\ref{fig14} refer to eigenstates at nonzero energy, while the lower two rows refer to zero modes.
We observe that the latter are indeed confined into the sublattice A, as expected from Eq.~(\ref{def_alphaB_zero}).
Moreover, even the structure of energy eigenstates at nonzero energy, in particular of those that are close to zero modes, appears to be nontrivial, at least for the system sizes under consideration.
Investigating the evolution of the structure of these states with increasing the system size is an interesting problem for future research.

Next we consider a more general case in which the off-diagonal block of dimension $N_A \times N_B$ [bottom left block in Fig.~\ref{fig_sublattice}(b)] consists of $N_B - m$ columns that are linearly independent, while the remaining $m \geq 0$ columns are linear combinations of the $N_B - m$ linearly independent columns.
The rank of the matrix is given by Eq.~(\ref{def_rank_H_m}), i.e., ${\rm rank}(H) = V - (\delta N +2m)$, and the number of zero modes is ${\cal M} = \delta N + 2m$.
We show below that in this case, the wavefunctions of $\delta N + m$ zero modes are confined into the larger sublattice, i.e., sublattice A, while the wavefunction of the remaining $m$ zero modes are confined into the sublattice B.

We first rewrite Eqs.~(\ref{eq_columns_1}) and~(\ref{eq_columns}) at $m>0$ as
\begin{equation} \label{def_H_alphaB_m}
    \hat{H}|\alpha_B\rangle = \sum_{i = 1}^{N_B}v_i |s_i\rangle = \sum_{i = 1}^{N_B-m}v_i |s_i\rangle + \sum_{i = N_B-m+1}^{N_B}v_i |s_i\rangle \,
\end{equation}
and express the vectors that are not linearly independent as
\begin{equation}
    \label{eq_lindependent}
    |s_i\rangle = \sum_{j = 1}^{N_B - m} k_j^{(i)} |s_j\rangle\;, \quad i = N_B-m+1, \dots, N_B \,.
\end{equation}
This enables us to rewrite Eq.~(\ref{def_H_alphaB_m}) as 
\begin{align}
\label{eq_eq_columns2}
    \hat{H}|\alpha_B\rangle & = \sum_{i = 1}^{N_B-m}v_i |s_i\rangle + \sum_{i = N_B-m+1}^{N_B}v_i \sum_{j = 1}^{N_B - m} k_j^{(i)} |s_j\rangle  \\
    &= \sum_{i = 1}^{N_B-m}v_i |s_i\rangle + \sum_{j = 1}^{N_B - m} \omega_j |s_j\rangle 
    = \sum_{i = 1}^{N_B - m} (v_i + \omega_i) |s_i\rangle \,, \nonumber
\end{align}
where the coefficients $\omega_j$ are $\omega_j = \sum_{i = N_B-m+1}^{N_B}v_i k_j^{(i)}$.
As in Eq.~(\ref{eq_columns_1}) one has $\hat H |\alpha_B\rangle = 0$, which yields $N_B - m$ conditions for the coefficients $v_i$,
\begin{equation}
\label{eq_zerocondition}
    v_i = -\omega_i = -\sum_{l = N_B-m+1}^{N_B}v_l k_i^{(l)}\;, \quad i = 1, \dots, N_B-m \,.
\end{equation}
This suggests that one is free to choose $m$ coefficients from subsystem B for any zero mode, while the remaining $N_B - m$ coefficients are set by the condition Eq.~(\ref{eq_zerocondition}).
In other words, one may view the effective dimensionality of the state $|\alpha_B\rangle$ to be given by $m$.

Since the Hamiltonian matrix is symmetric, there are again $N_B - m$ linearly independent rows (and therefore columns) in the off-diagonal block of size $N_B \times N_A$ [upper right block in Fig.~\ref{fig_sublattice}(b)], while the other $N_A - N_B + m = \delta N + m$ are linear combinations of the first $N_B - m$ ones.
Therefore one can similarly show that one is free to choose $\delta N + m$ coefficients from subsystem A for any zero mode, while the other $N_B - m$ are set by a condition similar to Eq.~(\ref{eq_zerocondition}).
One may hence view the effective dimensionality of the state $|\alpha_A\rangle$ to be given by $\delta N + m$.

In summary, one can consider a general form of zero modes that is a superposition of $m$ states from subsystem B and $\delta N + m$ states from subsystem A.
Since it is possible to construct symmetrized zero modes such that they are confined into a single subsystem only, see Eq.~(\ref{def_symmetrized_zeromodes}), this gives rise to $m$ zero modes that are confined into subsystem B and the other $\delta N + m$ zero modes to be confined into subsystem A.

This case is relevant for the description of zero modes in the billiard B2.
In Fig.~\ref{fig13} we show examples of the wavefunction amplitudes in this billiard at $V=1534$.
The upper three rows in Fig.~\ref{fig13} refer to eigenstates at nonzero energy, while the lower row refers to zero modes.

We note that the zero modes obtained after numerically diagonalizing the Hamiltonian matrix (such as those shown in the lower row of Fig.~\ref{fig13}) do not necessary belong in a single sublattice only.
We verified that if one diagonalizes the subspace of all zero modes according to the  operator $\hat \Gamma$ from Eq.~(\ref{def_gamma}), one indeed obtains new orthogonal zero modes that are eigenstates of $\hat\Gamma$ with eigenvalues $\pm 1$.
Furthermore, $m$ of these new zero modes are confined into subsystem B and the other $\delta N + m$ are confined into subsystem A.

\section{Protocols for lattice bipartitions} \label{app:bipartitions}

In Sec.~\ref{sec:entanglement} we studied the bipartite entanglement entropies of many-body Hamiltonian eigenstates in tight-binding billiards.
In each billiard we considered four bipartitions as sketched in Fig.~\ref{fig3}.
The protocols for performing these bipartitions are given below.
The procedure is terminated once the number of lattice sites in subsystem $\cal A$ is equal to the number of lattice sites in its complement. \\
(1) Bipartitions $\mathrm{a}_1$ and $\mathrm{b}_1$ (horizontal bipartition):
a subsystem $\cal A$ is constructed by gradually adding rows from bottom to top.
In each row individual sites are added from left to right. \\
(2) Bipartitions $\mathrm{a}_2$ and $\mathrm{b}_2$ (vertical bipartition):
a subsystem $\cal A$ is constructed by gradually adding columns from left to right.
In each column individual sites are added from bottom to top. \\
(3) Bipartitions $\mathrm{a}_3$ and $\mathrm{b}_3$ (diagonal bipartition):
a subsystem $\cal A$ is constructed by first choosing a particular corner site of a lattice, and then gradually adding lattice sites along the diagonals that are closest to that site.
In billiard 1 ($\mathrm{a}_3$) we start at the bottom right corner site and add lattice sites along the diagonals, such that each of the diagonals runs from top right to bottom left.
In billiard 2 ($\mathrm{b}_3$) we start in the bottom left corner site and add
lattice sites along the diagonals, such that each of the diagonals runs from top left to bottom right. \\
(4) Bipartitions $\mathrm{a}_4$ and $\mathrm{b}_4$ (symmetric bipartition):
a subsystem $\cal A$ is constructed by first drawing a line that connects two edge points of billiards in continuum and then gradually rotate the line by increasing the angle of the line with the horizontal axis.
Lattice sites of tight-binding billiards that are located in the region between the original line and the rotated line belong to subsystem $\cal A$.
In billiard 1 ($\mathrm{a}_4$) we draw a line that connects the points $(0,0)$ and $(L,0)$, see Fig.~\ref{fig1}(a) and the definition of the billiard in Eq.~(\ref{eq:Sinai}).
We then increase the angle of the line with the lower horizontal line in $L$ steps, such that in the final step the line connects the points $(0,0)$ and $(L,L)$.
In each step we add new lattice sites by rows, from bottom to top and from left to right.
In billiard 2 ($\mathrm{b}_4$) we draw a line that connects the points $(L, L)$ and $(0, y_h - R_h)$, see Fig.~\ref{fig1}(c) and the definition of the billiard in Eq.~(\ref{eq:Loka}).
We then increase the angle of the line with the upper horizontal line in $L$ steps, such that in the final step the line connects the points $(L,L)$ and $(x_v - R_v, 0)$.
In each step we add new lattice sites by rows, from top to bottom and from right to left.

\section{Calculation of entanglement entropies} \label{app:entropies}

In the calculations of eigenstate entanglement entropies, we first randomly select a many-body eigenstate $|m\rangle$.
Denoting the single-particle energy eigenstates as $\{ |q\rangle = \hat c_q^\dagger |\emptyset\rangle;\, q = 1, ..., V \}$, we randomly choose (with probability 1/2) for each $|q\rangle$ whether it is occupied or empty.
Then, the many-body eigenstate is constructed as $|m\rangle = \prod_{\{q_l\}_m} \hat c_{q_l}^\dagger |\emptyset\rangle$, where $\{ q_l\}_m$ represents the set of occupied single-particle energy eigenstates in $|m\rangle$.

Next we construct a one-body correlation matrix~\cite{peschel_03} of $|m\rangle$ with matrix elements
\begin{equation} \label{def_rho_m}
(\rho_m)_{ij} = \bra{m}\hat c_i^\dagger \hat c_j \ket{m} \,,
\end{equation}
where $i,j,=1,...,V$.
The matrix elements can be calculated using a unitary transformation between the operators $\hat c_i^\dagger, \hat c_i$ that create and annihilate a particle on lattice site $i$, and the operators $\hat c_q^\dagger, \hat c_q$ that create and annihilate a particle in an energy eigenstate $|q\rangle$.

We determine a lattice bipartition into the subsystem ${\cal A}$ with $V_{\cal A}$ sites and the subsystem ${\cal B}$ with $V_{\cal B}$ sites, where $V = V_{\cal A} + V_{\cal B}$ and $V_{\cal A} \leq V_{\cal B}$.
The one-body correlation matrix $\rho_m$ from Eq.~(\ref{def_rho_m}) is then restricted to entries $i,j$ from the subsystem ${\cal A}$, and the corresponding eigenvalues are denoted $\{\Lambda_i\}$, $i=1,...,V_{\cal A}$.
The von Neumann entanglement entropy can be obtained as~\cite{peschel_eisler_09}
\begin{equation}
S_m = \sum_{i=1}^{V_{\cal A}} -\left[\Lambda_i \ln\Lambda_i + (1-\Lambda_i) \ln(1-\Lambda_i) \right] \,.
\end{equation}
Similarly, the 2nd R\' enyi entanglement entropy is given by~\cite{hackl_bianchi_21}
\begin{equation} \label{S2_m}
S_{m}^{(2)} = \sum_{i=1}^{V_{\cal A}} - \ln\left[\Lambda_i^2 + \left(1-\Lambda_i\right)^2\right] \,.
\end{equation}
The average eigenstate entanglement entropies $\overline{S}$ and $\overline{S^{(2)}}$ are then calculated according to Eq.~(\ref{def_Saverage}), i.e., by averaging over contributions from different randomly selected many-body eigenstates $|m\rangle$.

\section{Closed-form expression for the 2nd R\' enyi entanglement entropy} \label{app:renyi}

The main input for the derivation is the distribution ${\cal F}_f(x)$ of eigenvalues of the one-body correlation matrix $\rho_m$, restricted to entries $i,j$ from the subsystem ${\cal A}$.
Here $f$ denotes the subsystem fraction $f = V_{\cal A}/V \in (0,1/2]$.
As pointed out in~\cite{liu_chen_18}, the restricted $\rho_m$ of a typical many-body eigenstate $|m\rangle$ of random quadratic Hamiltonians such as the Dirac SYK2 model belongs to the $\beta$-Jacobi ensemble with $\beta=2$.
Then, the distribution of the corresponding eigenvalues of systems at half-filling is~\cite{liu_chen_18}
\begin{equation}
    {\cal F}_f(x) = \frac{1}{2\pi f} \frac{\sqrt{(f_+-x)(x-f_-)}}{x(1-x)} 1_{[f_-,f_+]} \;,
\end{equation}
where $1_{[f_-,f_+]}$ indicates that the distribution is nonzero for $f \in [f_-, f_+]$, with $f_\pm = \frac{1}{2}(1 \pm 2\sqrt{f(1-f)})$.
The average entanglement entropy of $n$-th R\' enyi entropy is
$\overline{{\cal S}^{(n)}} = V_{\cal A} \int {\cal F}_f(x) S^{(n)}(x) dx$, where $S^{(n)}(x) = \frac{1}{1-n} \ln[x^n + (1-x)^n]$.

We next perform two steps that are identical to those in~\cite{zhang_liu_20}.
(a) We express
\begin{equation} \label{simplify_Sn_1}
    \ln[x^n + (1-x)^n] = \sum_{j=0}^{n-1} \ln[x- \xi_j(1-x)] \;,
\end{equation}
where $\xi_j = e^{i\pi(2j+1)/n}$ for $j=0,1,...,n-1$.
(b) We further express each summand on the r.h.s.~of Eq.~(\ref{simplify_Sn_1}) as
\begin{align} \label{simplify_Sn_2}
    \ln[x - \xi_j(1-x)] & = \ln[x(1+\xi_j) - \xi_j] \nonumber \\
    & = \ln\Bigg[(1+\xi_j)\left(x - \frac{\xi_j}{1+\xi_j}\right)\Bigg] \nonumber \\
    & = \ln(1+\xi_j) + \ln(x-d_j) \;,
\end{align}
where $d_j = \xi_j/(1+\xi_j)$.
Note that these steps assume that $n$ is a positive even integer and that the contributions to $\xi_j$ come in pairs $(j', n-1-j')$, thereby guaranteeing the sum in Eq.~(\ref{simplify_Sn_1}) to be real.
Equations~(\ref{simplify_Sn_1}) and~(\ref{simplify_Sn_2}) enable us to express the average entanglement entropy $\overline{{\cal S}^{(n)}}$ as a sum of two contributions,
\begin{equation} \label{def_Sn1}
    {\cal S}_{n_a} = \frac{1}{1-n} \frac{1}{2\pi f} \sum_{j=1}^{n-1} \ln(1+\xi_j)
    \int_{f_-}^{f_+} \!\!\! \frac{\sqrt{(f_+ - x)(x - f_-)}}{x(1-x)} dx \;
\end{equation}
and
\begin{equation} \label{def_Sn2}
    {\cal S}_{n_b} = \frac{1}{1-n} \frac{1}{2\pi f} \sum_{j=1}^{n-1}
    \int_{f_-}^{f_+} \!\!\! \frac{\sqrt{(f_+ - x)(x - f_-)}}{x(1-x)} \ln(x-d_j) dx \,,
\end{equation}
such that $\overline{{\cal S}^{(n)}} = V_{\cal A}({\cal S}_{n_a} + {\cal S}_{n_b})$.

We now focus on the 2nd R\' enyi entropy by setting $n=2$.
In this case, $\sum_{j=0}^1 \ln(1+\xi_j) = \ln 2$ and hence ${\cal S}_{2_a} \equiv {\cal S}_{a}$ in Eq.~(\ref{def_Sn1}) can be simplified to
\begin{align}
    S_{a} & = - \frac{1}{2\pi f}  \int_{f_-}^{f_+} \!\!\! \frac{\sqrt{(f_+ - x)(x - f_-)}}{x(1-x)} dx \, \ln 2 \nonumber \\
    & = - \left(\frac{1 - 2 \sqrt{f_+ f_-}}{2f} \right) \ln 2 \nonumber \\
    & = - \left( \frac{1 - \sqrt{1-4f(1-f)}}{2f} \right) \ln 2\;,
\end{align}
where in the last step we used the relation $2\sqrt{f_+ f_-} = \sqrt{1-4f(1-f)}$.
Evaluation of ${\cal S}_{2_b}$ in Eq.~(\ref{def_Sn2}) is however more tedious.
Defining
\begin{equation} \label{def_I1}
    I_1(x,f_+,f_-) = \int \frac{\sqrt{(f_+ - x)(x - f_-)}}{x(1-x)} dx
\end{equation}
one can solve it per-parts, i.e., expressing ${\cal S}_{2_b} = {\cal S}_{b1} + {\cal S}_{b2}$, where
\begin{equation} \label{def_Sb1}
    {\cal S}_{b1} = - \frac{1}{2\pi f} \sum_{j=0}^1
    \left( \left. I_1(x,f_+,f_-) \ln(x-d_j)\right|_{f_-}^{f^+} \right)
\end{equation}
and
\begin{equation} \label{def_Sb2}
    {\cal S}_{b2} = \frac{1}{2\pi f} \sum_{j=0}^1
    \left( \int_{f_-}^{f^+} I_1(x,f_+,f_-) \frac{1}{x-d_j} dx \right) \;.
\end{equation}
The coefficients $d_j$ in Eqs.~(\ref{def_Sb1}) and~(\ref{def_Sb2}) are $d_0 = \xi/(1+\xi)$ and $d_1 = \xi^*/(1+\xi^*)$, with $\xi=i$.
The integral $I_1(x,f_+,f_-)$ from Eq.~(\ref{def_I1}) was already computed in~\cite{zhang_liu_20} and it yields (omitting its dependence on $f_+$, $f_-$ for clarity)
\begin{align}
    I_1(x) & = -2 \arctan F(x) +
    2 \sqrt{f_+ f_-} \arctan \left( \frac{\sqrt{f_+ f_-}}{f_+} F(x) \right)
    \nonumber \\
    & + 2 \sqrt{f_+ f_-} \arctan \left( \frac{\sqrt{f_+ f_-}}{f_-} F(x) \right) \;,
    \label{def_I1_b}
\end{align}
where $F(x) = \sqrt{\frac{f_+ - x}{x - f_-}}$.
One observes that
$I_1(x=f_+) = 0$ and
$I_1(x=f_-) = - \pi + 2\pi \sqrt{f_+ f_-}$, which gives rise to a rather simple expression for ${\cal S}_{b1}$ from Eq.~(\ref{def_Sb1}),
\begin{align}
    {\cal S}_{b1} & = - \frac{1 - 2\sqrt{f_+ f_-}}{2f} \sum_{j=0}^1 \ln(f_- - d_j) \\
    & = \frac{1 - \sqrt{1-4f(1-f)}}{2f} \Big(2 \ln 2 - \ln\left[1+4f(1-f)\right] \Big) \,. \nonumber
\end{align}
On the other hand, evaluation of ${\cal S}_{b2}$ from Eq.~(\ref{def_Sb2}) requires to solve integrals of the form
\begin{align} \label{def_I2}
    I_2 & = \int_{f_-}^{f_+} \arctan \left( \eta \sqrt{\frac{f_+ - x}{x - f_-}} \right)
    \frac{1}{x-d_j} dx \nonumber \\
    & = \pi \ln \left( \frac{1 + \eta\sqrt{\frac{f_+ - d_j}{f_- - d_j}}}{1+\eta} \right) \;,
\end{align}
where it follows from Eq.~(\ref{def_I1_b}) that the values of $\eta$ are:
$\eta=1$, $\eta=\sqrt{f_-/f_+}$ and $\eta = \sqrt{f_+/f_-}$.
This gives rise to three contributions to ${\cal S}_{b2}$,
which we express as ${\cal S}_{b2} = {\cal S}_{b2a} + {\cal S}_{b2b} + {\cal S}_{b2c}$.
The first contribution, ${\cal S}_{b2a}$, stems from the first term on the r.h.s.~of Eq.~(\ref{def_I1_b}), i.e., it corresponds to $\eta=1$ in Eq.~(\ref{def_I2}), and it gives
\begin{align}
    {\cal S}_{b2a} & = \frac{1}{f} \ln 2 + \frac{1}{2f} \ln\left[ 1 +4f(1-f) \right] \nonumber \\
    & - \frac{1}{f} \ln\left( 1 + \sqrt{1+4f(1-f)} \right)\;.
\end{align}
The second contribution, ${\cal S}_{b2b}$, stems from the second term on the r.h.s.~of Eq.~(\ref{def_I1_b}), i.e., it corresponds to $\eta=\sqrt{f_-/f_+}$ in Eq.~(\ref{def_I2}), and it gives
\begin{align}
    {\cal S}_{b2b} & = \frac{\sqrt{1-4f(1-f)}}{2f}
    \Big[ - \frac{1}{2} \ln[1+4f(1-f)] \nonumber \\
    &  + \ln\left( \sqrt{1+4f(1-f)} + \sqrt{1-4f(1-f)} \right) \nonumber \\
    &  - \ln \left( 1+ \sqrt{1-4f(1-f)} \right) \Big]\;.
\end{align}
The third contribution, ${\cal S}_{b2c}$, stems from the third term on the r.h.s.~of Eq.~(\ref{def_I1_b}), i.e., it corresponds to $\eta=\sqrt{f_+/f_-}$ in Eq.~(\ref{def_I2}), and it is equal to the second contribution, i.e.,
\begin{equation}
    {\cal S}_{b2c} = {\cal S}_{b2b} \;.
\end{equation}
All together, we get
$\overline{{\cal S}^{(2)}} = V_{\cal A}({\cal S}_{a} + {\cal S}_{b1} + {\cal S}_{b2a} + 2{\cal S}_{b2b})$, which is
\begin{align}
    &\overline{{\cal S}^{(2)}} = V_{\cal A} \frac{\ln 2}{2f} \Bigg[ 1 - \sqrt{1-u(f)} - 2 \log_2 \left(\frac{1+\sqrt{1+u(f)}}{2}\right) \nonumber \\
    & - 2 \sqrt{1-u(f)} \log_2 \left(\frac{1+\sqrt{1-u(f)}}{\sqrt{1+u(f)}+\sqrt{1-u(f)}}\right) \Bigg] \;,
\end{align}
where $u(f) = 4f(1-f)$.
Finally, we observe that by replacing $\sqrt{1-u(f)} \to 1-2f$, which is strictly speaking valid only for $f \leq 1/2$, we get Eq.~(\ref{def_S2_rmt}) in the main text, which is symmetric w.r.t.~$f=1/2$, i.e., it is applicable at any $0<f<1$.

\section{Translationally-invariant free fermions on a square lattice with regular boundaries} \label{app:Sfree}

\begin{figure}[!t]
\centering
\includegraphics[width=0.98\columnwidth]{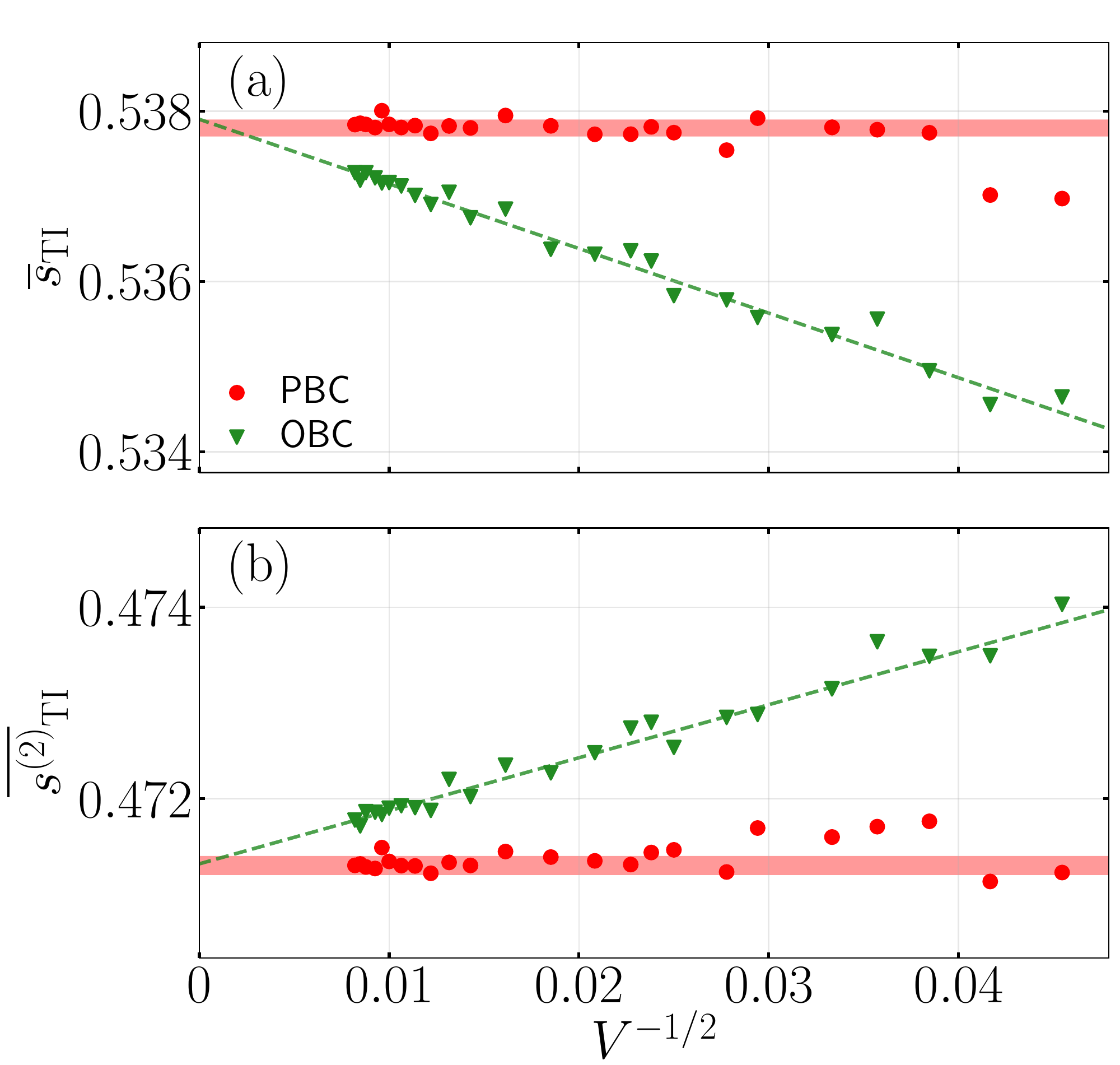}
\caption{
Volume-law coefficients $\overline{s}_{\rm TI}$ and $\overline{s^{(2)}}_{\rm TI}$ of the average eigenstate entanglement entropies for TI free fermions in 2D lattices with regular boundaries, using PBC (circles) and OBC (triangles).
(a) Results for von Neumann entanglement entropy, (b) results for the 2nd R\' enyi entropy, both at the subsystem fraction $f=1/2$.
The volume-law coefficients are defined using Eqs.~(\ref{def_volume_coef}) and~(\ref{def_Saverage}), where at each $V$ we average over $M = 5000$ many-body eigenstates.
Horizontal red lines are results for the same model in 1D and 3D lattices, (a) $\overline{s}_{\rm TI} = 0.5378$~\cite{vidmar_hackl_17} and (b) $\overline{s^{(2)}}_{\rm TI} = 0.4713$~\cite{lydzba_rigol_21}.
The width of the lines correspond to the estimated error bar $10^{-4}$.
Dashed lines are linear fits to the OBC results at $V \geq 3000$.
}\label{fig12}
\end{figure}

We complement the results for the average eigenstate entanglement entropies of tight-binding billiards in Sec.~\ref{sec:entanglement} with those for translationally invariant (TI) free fermions on square lattices with regular (square) boundaries, using either periodic or open boundary conditions (PBC or OBC, respectively).
It was suggested~\cite{vidmar_hackl_17} that the volume-law contribution to the average eigenstate entanglement entropies of TI free fermions on hypercubic lattices with regular boundaries may be independent of dimensionality.
However, the numerical calculations were so far only reported in 1D~\cite{vidmar_hackl_17, lydzba_rigol_20} and 3D lattices~\cite{lydzba_rigol_21} for the von Neumann entropy, and in 1D and 3D lattices for the 2nd R\' enyi entropy~\cite{lydzba_rigol_21}.

Previous numerical studies in 1D and 3D lattices determined the volume-law coefficients at the subsystem fraction $f=1/2$ to be $\overline{s}_{\rm TI} = 0.5378$ for the von Neumann entropy~\cite{vidmar_hackl_17}, and $\overline{s^{(2)}}_{\rm TI} = 0.4713$ for the 2nd R\' enyi entropy~\cite{lydzba_rigol_21}.
Our numerical results for the 2D lattices at different system sizes $V$ are shown as symbols in Fig.~\ref{fig12}.
The horizontal lines in Fig.~\ref{fig12} denote the predictions $\overline{s}_{\rm TI}$ and $\overline{s^{(2)}}_{\rm TI}$ from 1D and 3D lattices given above, and the width of the horizontal lines correspond to an estimated error bar $10^{-4}$.
One observes that the 2D results using PBC agree with the 1D and 3D predictions within the error bar already for moderately large systems.
This is expected since the finite-size corrections to the average eigenstate entanglement entropies of TI free fermions with PBC are small~\cite{vidmar_hackl_17, bianchi_hackl_22}.
On the other hand, the 2D results using OBC exhibit larger finite-size corrections, which were also observed in 1D systems~\cite{lydzba_rigol_20}.
Still, a simple extrapolation of the results at large V, see the dashed lines in Fig.~\ref{fig12}, suggest that the OBC results agree with the PBC results to high precision, and will eventually become identical in the thermodynamic limit.
It is interesting to note that the finite-size corrections to both $\overline{s}_{\rm TI}$ and $\overline{s^{(2)}}_{\rm TI}$ appear to scale as $V^{-1/2} = L^{-1}$, where $L$ is the linear size of the system.
The same scaling with $L$ was observed for $\overline{s}_{\rm TI}$ in one dimension~\cite{lydzba_rigol_20}.

In conclusion, our numerical results for the volume-law coefficients in 2D TI free fermion models are consistent with results in the same model in dimensions one and three, and firmly establish that these volume-law coefficients are in asymptotically large systems different from those found in quantum-chaotic quadratic systems, which are obtained from Eqs.~(\ref{def_S_rmt}) and~(\ref{def_S2_rmt}).

\bibliographystyle{biblev1}
\bibliography{references}

\end{document}